\documentclass[11pt]{article}
\usepackage{amsmath, amssymb, amscd, amsthm, amsfonts}
\usepackage{graphicx}
\usepackage{hyperref}
\usepackage{here}

\usepackage[round]{natbib}
\usepackage{bm}
\usepackage{xcolor}

\usepackage[T1]{fontenc}
\usepackage[utf8]{inputenc}
\usepackage{authblk}

\newtheorem{thm}{Theorem}[]
\newtheorem{lem}{Lemma}[]
\newtheorem{cor}{Corollary}[]
\newtheorem{prop}{Proposition}[]

\usepackage{amsmath}               
  {
      \theoremstyle{plain}
      \newtheorem{assumption}{Assumption}
  }

\newcommand{\argmax}{\mathop{\rm arg~max}\limits}

\newcommand\real{\mathbb{R}}

\newcommand{\1}{\mathbf{1}}

\newcommand{\A}{\mathcal{A}}

\newcommand{\T}{\mathcal{T}}

\newcommand{\Polya}{P\'{o}lya\ }
\newcommand{\F}{\mathbf{F}}

\newcommand{\G}{\mathbf{G}}

\newcommand{\cP}{\mathcal{P}}

\newcommand{\cG}{\mathcal{G}}
\newcommand{\bG}{\bm{\mathcal{G}}}

\begin{document}

\title{Unsupervised Tree Boosting for Learning Probability Distributions}

\author[1]{Naoki Awaya\thanks{nawaya@stanford.edu}}
\author[2]{Li Ma \thanks{li.ma@duke.edu}}
\affil[1]{Department of Statistics, Stanford University}
\affil[2]{Department of Statistical Science, Duke University}

\date{\today}

\maketitle

\begin{abstract}
We propose an unsupervised tree boosting algorithm for inferring the underlying sampling distribution of an i.i.d.\ sample based on fitting additive tree ensembles in a fashion analogous to supervised tree boosting. Integral to the algorithm is a new notion of ``addition'' on probability distributions that leads to a coherent notion of ``residualization'', i.e., subtracting a probability distribution from an observation to remove the distributional structure from the sampling distribution of the latter. We show that these notions arise naturally for univariate distributions through cumulative distribution function (CDF) transforms and compositions due to several ``group-like'' properties of univariate CDFs. While the traditional multivariate CDF does not preserve these properties, a new definition of multivariate CDF can restore these properties, thereby allowing the notions of ``addition'' and ``residualization'' to be formulated for multivariate settings as well. This then gives rise to the unsupervised boosting algorithm based on forward-stagewise fitting of an additive tree ensemble, which sequentially reduces the Kullback-Leibler divergence from the truth. The algorithm allows analytic evaluation of the fitted density and outputs a generative model that can be readily sampled from. We enhance the algorithm with scale-dependent shrinkage and a two-stage strategy that separately fits the marginals and the copula. The algorithm then performs competitively to state-of-the-art deep-learning approaches in multivariate density estimation on multiple benchmark data sets.
\end{abstract}

\section{Introduction}
\label{sec: Introduction}
In supervised learning such as classification and regression, boosting is acknowledged as one of the most powerful algorithms. It is acclaimed for the ability to overcome the curse of dimensionality and achieve a desirable balance in bias-variance trade-off. The most popular boosting algorithms can be thought of as sequentially fitting an additive ensemble of weak learners, often in the form of regression or classification trees \cite[e.g.,][]{friedman2001greedy, hastie2009elements}.
The success of tree boosting in supervised problems suggests that a similar strategy might also prevail in unsupervised problems, where the ultimate objective involves learning the structures of some unknown probability distribution based on a collection of training data from that distribution.

Our aim in this paper is to formulate a new additive tree model framework for probability distributions along with an unsupervised boosting algorithm that inherits the strength of the supervised boosting.
Our approach is motivated by the observation that some highly effective supervised tree boosting algorithms are fit in each iteration based on a set of residuals rather than the original observations, thereby substantially simplifying the optimization task in each iteration. 
To realize this strategy in the unsupervised context, we introduce a notion of {\em addition} specialized for probability measures which leads to a natural concept of the {\em residual} of an observation after ``subtracting'' a probability measure from it. 
It should be clarified that in the unsupervised setting, the natural operation of addition---such as through taking weighted averages---does not work, as it is not straightforward to find such an embedding that renders a conceptually and computationally simple notion of ``residualization'' of an observation, which removes a fitted measure from the underlying sampling distribution.

The notions of ``addition'' and ``residualization'' for probability distributions are formulated in terms of cumulative distribution function (CDF) transforms and compositions. 
We start from the case of univariate distributions, for which the addition of two measures can be defined simply in terms of a composition of their CDFs whereas the residual of an observation from subtracting a measure is simply the application of the corresponding CDF transform to that observation. 
In generalizing this notion of addition to multivariate distributions on $\real^d$ with $d>1$, however, the classical notion of the multivariate CDF, which maps $\real^d$ to the interval $(0,1]$ is unsatisfactory, as easily seen, for example, by the fact that one can neither define a composition of two such CDFs nor define residuals that still lie in $\real^d$. More fundamentally, multivariate CDFs do not preserve a set of ``group-like'' properties of 1D CDFs that underlie the notion of addition and residualization.  
Interestingly, it can be shown that a proper notion of the CDF for multivariate distributions, which maps $\real^d$ to $\real^d$, does exist for probability measures defined on tree-partition structures (tree measures) and it naturally generalizes the notions of addition and residuals to multivariate settings. 

Based on these notions, we introduce an unsupervised tree boosting algorithm for learning probability measures based on forward-stagewise (FS) fitting of an additive tree ensemble. Our algorithm in each iteration completes two operations that resemble those in supervised boosting: (i) computing the current residuals by subtracting the fitted measure at the current iteration from the observations and (ii) fitting a tree-based weak learner on the residuals and adding the estimated distribution to the current fit. The algorithm enables straightforward analytical evaluation of the probability density of the fitted distribution and produces a generative model for the fitted measure that can be directly sampled from.

{
Because the notion of addition on probability measures in our boosting framework takes the form of function compositions, one can also view our approach as a normalizing flow (NF) \citep[see i.e.,][]{papamakarios2019neural, papamakarios2021normalizing, kobyzev2020normalizing}. In NF approaches, one seeks to find a sequence of transformations that moves the observed distribution into a baseline distribution such as the uniform and the Gaussian.
Most notably, \cite{inouye2018deep} introduced a tree-based NF method through composing a class of transforms that are equivalent to the generalized CDF transforms we introduce. As such, their NF algorithm---called ``tree density destructors'' is in essence the same as our boosting algorithm aside from the difference in the choice of the base learner, the strategies in regularization, and the other boosting-inspired specification strategies. These different practical choices we make---which are largely motivated from the boosting perspective---do lead to substantial differences in empirical performance and computational efficiency, as we will demonstrate in our numerical experiments.
Besides, formulating the algorithm from the boosting perspective also allows us to provide a more rigorous theoretical grounding for the algorithm. Therefore, our main contribution is not in the novelty of the algorithm itself, but in connecting boosting and NF in both theory and practice.
In summary, our contributions are:
\begin{enumerate}
    \item {\it Reformulation of tree-based NFs as tree boosting.} We introduce a formal notion of addition on probability measures that leads to be group structure on (generalized) CDF transforms, based on which we show that tree-based NFs can be understood as iterative fitting of an additive ensemble of probability measures in ways analogous to iterative fitting of weak learners to residuals under classical tree boosting for supervised problems. 
    \item {\it Theoretical justification.} The boosting formulation allows us to justify the iterative algorithm formally from a decision-theoretic perspective in terms of sequential minimization of the Kullback-Leibler divergence between the model and the unknown true measure.  
    In addition, analogous to supervised tree boosting \citep{breiman2004population}, we show that the unsupervised boosting is ``highly expressive'' in the sense that a wide class of distributions can be represented or well approximated by a finite combination of highly constrained (or ``weak'') tree-based density models. 
    \item {\it Methodological improvement.}  The boosting formulation allows us to incorporate methodological techniques---originally developed for supervised problems.
    These include guidelines for choosing the number of trees, setting the appropriate level of shrinkage/regularization, and choosing/specifying the base learner. We provide a comprehensive empirical evaluation of our proposed boosting by comparing it with the density destructor \citep{inouye2018deep} and the state-of-the-art NF algorithms such as MAF \citep{papamakarios2017masked} using simulation examples and benchmark data sets.
    The results suggest that our new algorithm substantially improves performance over tree density destructors and is competitive to the state-of-the-art deep-learning based NF algorithms at a substantially small computation cost.  
    The decision-theoretic formulation also leads to a natural measure of variable importance based on the respective contribution of each dimension in reducing the overall KL divergence, which provides additional insights on the relevance of each dimension in characterizing the underlying distribution and allows effective variable screening.
\end{enumerate}
}

{
We note that boosting for unsupervised learning has been considered by \cite{ridgeway2002looking, rosset2002boosting, cui2021gbht}. 
These previous attempts however aim at constructing an ensemble in the form of a weighted average of probability measures and fit such ensembles through gradient boosting \citep{mason1999boosting} under various loss functions.
These unsupervised gradient boosting algorithms have yet to be demonstrated to be computationally efficient or perform well in high-dimensional continuous sample spaces.
}

All proofs are given in  {Appendix~\ref{supp: proofs} and \ref{sec: Expressive power of the tree ensemble}}.

\section{Method}
\label{sec: method}
In this section, we start by defining notions of addition and residuals for one-dimensional settings and next generalize them for multivariate distributions to introduce the new tree boosting algorithm. 

\subsection{CDF-based Addition and Residualization for Univariate Distributions}
\label{subsec: CDF-based addition rule}
Without loss of generality, let $(0,1]$ represent the one-dimensional sample space. For ease of exploration, we shall assume that the distributions are absolutely continuous on the sample space with respect to the Lebesgue measure and have full support. 

We first make an observation that if a random variable $X\sim G$, its sampling distribution, then $\G(X)\sim {\rm Unif}(0,1]$, where $\G$ denotes the CDF of $G$. (Throughout we will use bold font letters to indicate the CDFs of the corresponding distributions.) As such, the CDF transform ``removes'' the distributional structure of $G$ from the sampling distribution of $X$.
Thus one can think of $r=\G(X)$ as a ``residual''. Moreover, Unif(0,1] serves as the notion of ``zero'', whose CDF is the identity map, in the space of probability distributions as it is the remaining distribution after ``subtracting'' the true sampling distribution $G$ from $X$.

Next we define a notion of ``adding'' two distributions $G$ and $H$ that is consistent with the above notion of ``subtraction'' or ``residualization''. Specifically, the addition of $G_1$ and $G_2$, denoted as ``$G_1\oplus G_2$'', should satisfy the property that if $X\sim G_1\oplus G_2$, then $r^{(1)}=\G_1(X)\sim G_2$. In other words, if the sampling distribution of $X$ is the ``sum'' of $G_1$ and $G_2$, then taking the ``residual'' of $X$ with respect to $G_1$ should result in a random variable distributed as $G_2$.
Such a notion of addition indeed exists:
\[
G_1\oplus G_2 \mathrm{\ is\ the\ distribution\ whose\ CDF\ is\ } \G_2 \circ \G_1
\]
where ``$\circ$'' denotes function composition. Note that this notion of addition is not commutative. That is, $G_1\oplus G_2\neq G_2\oplus G_1$. Fortunately, as we will see, the operation of fitting an additive ensemble as in supervised boosting requires only a non-abelian group-structure, which does not require the commutativeness of the underlying addition. As such, the loss of commutativeness will pose no difficulty in our construction of additive tree models and later an unsupervised boosting algorithm based on the new notions of addition and residuals. 

By iteratively applying such an addition, one can define the ``sum'' of $k(\geq 1)$ probability measures $G_1,\dots,G_k$.
Specifically, the sum of $G_1,\dots,G_k$, 
\[ G_1 \oplus \cdots \oplus G_k \mathrm{\ is\ the\ distribution\ whose\ CDF\ is\ }
        \G_k \circ \cdots \circ \G_1. \]
The following property provides the basis for sequential addition and residualization, analogous to those in supervised boosting.

\begin{prop}
\label{prop: residualizatoin for univariate cases}
If $G_1,\ldots,G_k$ have full support on $(0,1]$, (i.e., when the CDFs $\G_1,\dots,\G_k$ are strictly increasing), then for any $i=1,2,\ldots,k-1$
\begin{align*}
X \sim G_1 \oplus \cdots \oplus G_k
\quad \mathrm{\ if\ and\ only\ if} \quad 
r^{(i)} = \G_{i} \circ \cdots \circ \G_1(X)
    \sim
    G_{i+1}\oplus  \cdots \oplus G_k.
\end{align*}
\label{prop:seq_cdf}
\end{prop}
Proposition~\ref{prop:seq_cdf} then implies that such residualization can be applied sequentially. That is, if $r^{(k)}$ is the residual of $x$ after subtracting $G_1\oplus\cdots \oplus G_{k}$, then $r^{(0)}=x$, and for $k\geq 1$
\begin{align}
\label{eq:residuals}
r^{(k)} = \G_{k}(r^{(k-1)})=\G_{k}\circ \G_{k-1}(r^{(k-2)})\cdots = \G_{k} \circ \G_{k-1} \circ \cdots \circ \G_1(r^{(0)}).
\end{align}
The ``additivity'' induced by the composition of CDFs also induces an additivity on the corresponding log-likelihood. Specifically, suppose $g_i=dG_i/d\mu$ is the probability density function (pdf) of $G_i$ for $i=1,2,\ldots,k$ with respect to Lebesgue measure $\mu$. Then the density of the ensemble measure $F_k := G_1 \oplus \dots \oplus G_k$, $f_k=dF_k/d\mu$, satisfies
\[
f_k(x) = \prod_{i=1}^{k} g_i(r^{(i-1)}) \quad \text{or} \quad \log f_k(x) = \sum_{i=1}^{k} \log g_i(r^{(i-1)}).
\]

\begin{table}[t]
\centering
\begin{tabular}{lll}
& Boosting for regression & Boosting for probability measures \\ \hline
Addition & $h_1+\dots+h_k$              & $G_1 \oplus \cdots \oplus G_k$    \\
Residual & $y - \sum^{k-1}_{l=1}h_l(x)$ & $\G_{k-1} \circ \cdots \circ \G_1(x)$ \\
Zero     & $0$                          & Uniform distribution on $(0,1]^d$ 
\end{tabular}
\caption{Key concepts in boosting\label{table: notions}}
\end{table}

Table \ref{table: notions} summarizes the corresponding notions of addition, residuals, and zero in supervised boosting (in particular regression) and those in our unsupervised formulation.
With these new notions, we are ready to introduce a boosting algorithm for learning one-dimensional distributions. However, the more interesting application involves multivarate (in fact high-dimensional) distributions. As such, we first generalize these notions to multivariate cases, and then introduce a multivariate version of our boosting algorithm that contains the (less interesting) univariate scenario as a special case.

\subsection{Generalization to Multivariate Distributions}
\label{subsec: Generalizationfor multi-variate cases}
The above notions of addition and residuals do not find direct counterparts for multivariate measures if one uses the traditional definition of CDFs for multivariate distributions. In particular, because the traditional CDF is a mapping from $(0,1]^d$ to $(0,1]$ instead of $(0,1]^d$, we cannot even take the composition of the CDFs or compute the residuals, which should remain in the same space as the original observations. Beyond the minimal requirement that the appropriate notion of ``CDF'' should map from $(0,1]^d$ to $(0,1]^d$, it must also enjoy several group-like properties of univariate CDF's. 

We summarize four such properties that the ``CDF'' must satisfy to allow the definition of addition and residualization to carry over into the multivariate setting:
\begin{itemize}
    \item[(C1)] $\G$ is a mapping from $(0,1]^d$ to $(0,1]^d$.
    \vspace{-0.2em}
    
    \item[(C2)] $G$ is uniquely determined by $\G$.
    \vspace{-0.2em}
    
    \item[(C3)] If $X \sim G$, then $\G(X) \sim {\rm Unif}((0,1]^d)$, the ``zero''.
    \vspace{-0.2em}
    
    \item[(C4)] If $X\sim G_1\oplus G_2$, the distribution is  uniquely determined by its ``CDF'' $\G_2\circ \G_1$, then $\G_1(X) \sim G_2$.
\end{itemize}
\noindent Remark: (C1) and (C2) are needed for defining addition in terms of compositions. (C3) and (C4) are needed for the proper notion of residuals.

For the purpose of constructing a tree additive ensemble model and a boosting algorithm, one type of ``CDFs'' that satisfy these conditions are particularly useful as they are very easy to compute for probability distributions with piecewise constant densities defined on leafs of a recursive dyadic partitioning of the sample space. As one can imagine, efficient computation of the ``CDFs'' for tree-based models is critical as they will be computed many times during the fit to an additive ensemble.

\subsubsection{Characterizing probability measures on a recursive dyadic partition tree}
\label{subsubsec: Probability measure on a dyadic tree}

Next we describe the construction of this generalized notion of multivariate CDFs, which we call the ``tree-CDF'', due to its connection to recursive bifurcating partition trees. We start by introducing some additional notation related to recursive dyadic partitions.

A recursive dyadic partition of depth $R$ is a sequence of nested dyadic partitions \\ $\A^1,\A^2,\ldots,\A^{R}$ on the sample space $\Omega$. 
The first partition $\A^1$ only includes $\Omega$, and for $k=2,\dots,R$, the partition $\A^{k}$ consists of all the sets generated by dividing each $A \in \A^{k-1}$ into two children $A_l$ and $A_r$, where $A_l \cup A_r = A$ and $A_l \cap A_r = \emptyset$.
(Throughout, we use subscripts $l$ and $r$ to indicate left and right children respectively.) We can denote the recursive partition using a tree $T = \cup^{R}_{k=1} \A^k$. As such, we refer to the sets in the partitions as ``nodes''.
We call the collection of nodes in $\A^{R}$ the ``terminal'' nodes or ``leafs'' of $T$ and denote it by $\mathcal{L}(T)$; the nodes in other levels are the ``non-leaf'' nodes or ``interior'' nodes, which we denote by $\mathcal{N}(T)=T\backslash \mathcal{L}(T)$.

We consider partition trees with axis-aligned partition lines. In this case, a node $A \in T$ is of the following rectangular form
\begin{align}
    A =
    (a_1, b_1] \times \cdots \times (a_d, b_d].
    \label{A (rectangle)}
\end{align}
For a non-leaf node $A\in \mathcal{N}(T)$, the children $A_l$ and $A_r$ are generated by dividing $A$ in one of the $d$ dimensions, say $j^*$, 
\begin{align}
    A_l &=
    (a_1, b_1]
    \times 
    \cdots
    (a_{j^*}, c_{j^*}] \times
    \cdots
    (a_d, b_d] \quad \text{and} \quad
    A_r =
    (a_1, b_1]
    \times 
    \cdots
    (c_{j^*}, b_{j^*}] \times
    \cdots
    (a_d, b_d].
        \label{A_l and A_r (rectangle)}
\end{align}
In the following, for each partition tree $T$, we let $\cP_T$ be the class of probability measures that are conditionally uniform on the leafs of $T$ and have full support on $\Omega$. That is,
 \begin{align*}
     \cP_T =
     \left\{
        G : \text{$G$ has full support on $(0,1]^d$ and } G(\cdot \mid A) = \mu(\cdot \mid A) \text{ for every } A \in  \mathcal{L}(T)
     \right\},
 \end{align*}
 where $\mu$ is the uniform distribution, and $G(\cdot|A)$ and $\mu(\cdot|A)$ are the corresponding conditional distributions on $A$.
 
\begin{figure}[t]
\centering
\begin{tabular}{c}
    \includegraphics[height=4.7cm]{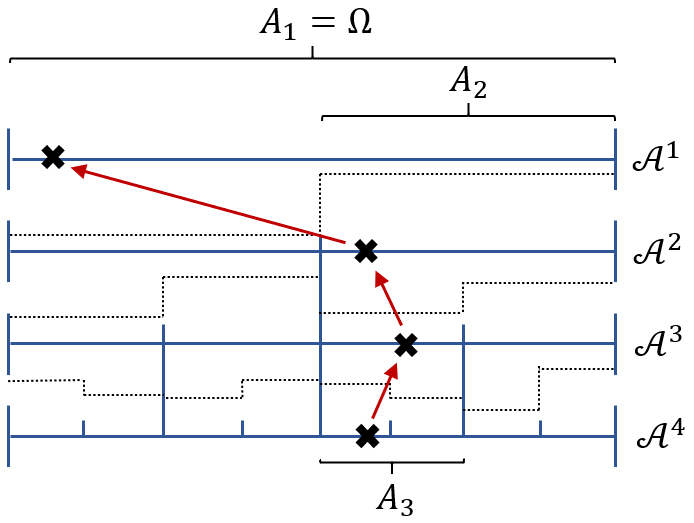}
\end{tabular}
\vspace{-0.5em}
\caption{
Visualization of the tree-based decomposition of a univariate CDF into three local moves.
The dotted lines indicate $G(A_l \mid A)/\mu(A_l \mid A)$ or $G(A_r \mid A)/\mu(A_r \mid A)$ on each node $A$.
}
\label{fig: 1D CDF decomposition}
\end{figure}
 
To generalize the CDF transform from univariate to multivariate cases, first we note an interesting multi-scale decomposition of the univariate CDF---a univariate CDF transform $\G$ for any distribution $G\in \cP_T$ on an observation $x$ can actually be computed sequentially in a fine-to-coarse fashion along the branch in the partition tree $T$ in which $x$ falls. Specifically, suppose $T$ has depth $R$, then for $x \in (0,1]$, let 
$\{A_k\}^{R}_{k=1}$ be a sequence of nodes in $T$ such that $A_k \in \A^k$ and 
 \begin{align*}
    x \in A_{R} \subset A_{R-1} \subset \cdots \subset A_1 = (0,1].    
 \end{align*}
The CDF transform $\G(x)$ can be decomposed into the composition of a sequence of ``local move functions''
\begin{align}
\label{eq:seq_cdf}
    \G(x) = \G_{A_1} \circ \cdots \circ \G_{A_{R-1}} (x),
\end{align}
where for any $A=(a,b]\in \mathcal{N}(T)$ with two children $A_l=(a,c]$ and $A_r=(c,b]$, the mapping $\G_{A}:A\rightarrow A$ is (up to a normalizing constant $\mu(A)$) the CDF of a dyadic piecewise constant density equal to $G(A_l|A)/\mu(A_l)$ on $A_l$ and $G(A_r|A)/\mu(A_r)$ on $A_r$. More precisely,  $\G_A:A\rightarrow A$ is given by
\[
\frac{\G_A(x)-a}{x-a} = \frac{G(A_l|A)}{\mu(A_l|A)} \quad \text{for $x\in A_l$}
\quad
\text{and}
\quad 
\frac{b-\G_A(x)}{b-x} = \frac{G(A_r|A)}{\mu(A_r|A)} \quad \text{for $x\in A_r$}.
\]
Note that the conditional measures and the input and output of $\G_A$ have the following relationship
\begin{align*}
    G(A_l|A)>\mu(A_l|A)
    &\quad \Leftrightarrow \quad
    G(A_r|A)<\mu(A_r|A)
    \quad \Leftrightarrow \quad
    \G_A(x)>x, \\
    G(A_l|A)<\mu(A_l|A)
    &\quad \Leftrightarrow \quad
    G(A_r|A)>\mu(A_r|A)
    \quad \Leftrightarrow \quad
    \G_A(x)< x.
\end{align*}
We call $\G_A$ a ``local move'' function because it moves a point in $A$ in the direction of the child node with less (conditional) probability mass than the (conditional) uniform measure as illustrated in Figure~\ref{fig: 1D CDF decomposition}. 
The amount of movement on $A$ is proportional to the probability mass differential between the two children of $A$ in $G$ relative to $\mu$.

If we think of applying the univariate CDF transform as ``subtracting'' the information contained in a probability measure from an observation, the decomposition in Equation~\ref{eq:seq_cdf} indicates that such subtraction can be done  sequentially through the local moves, each subtracting a piece of  information regarding the measure from the observation. This perspective leads to a generalization of the CDF transform for the multivariate case as we describe below.

\begin{figure}[tb]
\centering
\begin{tabular}{c}
    \includegraphics[height=4.0cm]{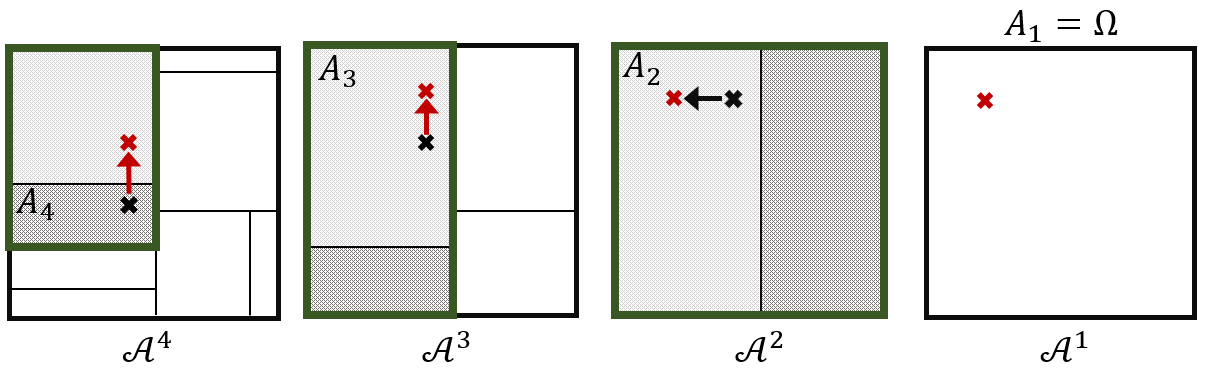}
    \\
    \includegraphics[height=4.25cm]{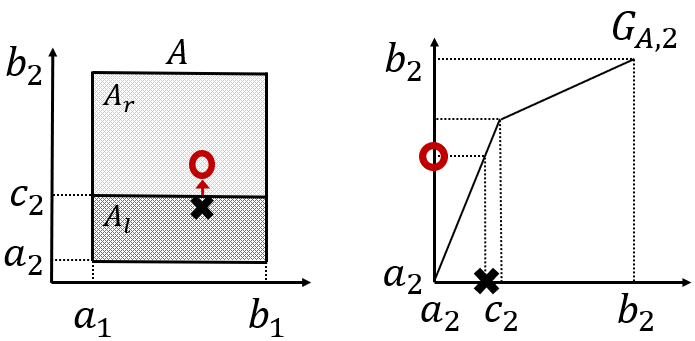}
\end{tabular}
\vspace{-0.5em}
\caption{
Top: Visualization of the local move functions in each level under $R=3$. The nodes with the darker color have higher conditional probabilities relative to the uniform measure.
Bottom: An example of $\G_A$ with $j^*=2$. The input and the output of $\G_A$ are indicated by $\times$ and $\circ$, respectively.
}
\label{fig: multi_scale CDF}
\end{figure}

For a point $x = (x_1,\dots,x_d) \in \Omega=[0,1)^d$, again let $T$ be a recursive dyadic partition tree of depth $R$ on the sample space $\Omega$, and $\{A_k\}_{k=1}^{R}$ the sequence of nodes in $T$ that contains $x$ as before. Then we define a mapping $\G:(0,1]^d\rightarrow (0,1]^d$ in terms of a sequence of fine-to-coarse local moves along that branch in $T$. Specifically, for a node $A\in\mathcal{N}(T)$ as in Equation~\ref{A (rectangle)} with children $A_l$ and $A_r$ attained from dividing $A$ in the $j^*$th dimension as described in Equation~\ref{A_l and A_r (rectangle)}, we define a local move mapping $\G_A:A\rightarrow A$ such that for any $x\in A$, $\G_A(x) = (\G_{A,1}(x),\dots,\G_{A,d}(x))$ where $\G_{A,j}(x)=x$ for all $j\neq j^*$, and
\[
\frac{\G_{A,j^*}(x)-a_{j^*}}{x_{j^*}-a_{j^*}} = \frac{G(A_l|A)}{\mu(A_l|A)} \quad \text{for $x\in A_l$}
\quad
\text{and}
\quad 
\frac{b_{j^*}-\G_{A,j^*}(x)}{b_{j^*}-x_{j^*}} = \frac{G(A_r|A)}{\mu(A_r|A)} \quad \text{for $x\in A_r$}.
\]
As illustrated in Figure~\ref{fig: multi_scale CDF}, similar to the univariate case, the local move mapping is nothing but (up to a normalizing constant $\mu(A)$) the CDF of a dyadic piecewise constant density on $A$, except that now in the multivariate setting there are a total of $d$ directions in which such a dyadic split can take place. As a transform, it moves $x$ in the direction of the child node with less probability mass relative to the uniform measure.

As before, we now define a mapping $\G:(0,1]^d\rightarrow (0,1]^d$, called a ``tree-CDF'', as the composition of these local move functions. That is, 
\begin{align*}
    \G(x) = \G_{A_1} \circ \cdots \circ \G_{A_{R-1}} (x).
\end{align*}
$\G$ is injective from $(0,1]^d$ to $(0,1]^d$ for any $G$ with full support on $(0,1]^d$. Additionally, because $\G_A$ is surjective for every $A$, $\G$ is also surjective.
One can also show that $\G$ is measurable.
Hence we have the following proposition that establishes Conditions (C1) for tree-CDFs, which is essential to defining the addition of multivariate distributions in terms of tree-CDF compositions.
\begin{prop}
\label{prop: G is bijective and measurable}
The tree-CDF mapping $\G: (0,1]^d \mapsto (0,1]^d$ is bijective and measurable for any $G\in \cP_T$.
\end{prop}

The next two theorems show that our construction of $\G$ satisfies Conditions (C2) and (C3) as well. That is, $\G$ uniquely determines $G$ and applying the $\G$ mapping to an observation effectively ``subtracts'' the distributional structure in $G$ from the sampling distribution of that observation.

\begin{thm}
    \label{thm: we can retrieve the information of a measure}
    A measure $G\in \cP_T$ for some partition tree $T$ can be determined by the tree-CDF mapping $\G$     as follows
    \begin{align*}
        G(B) = \mu(\{\G(x): x \in B\}) \quad \text{for all $B \in \mathcal{B}(\Omega)$.}
    \end{align*}
\end{thm}
\noindent 
Remark: 
{
Theorem~\ref{thm: we can retrieve the information of a measure} establishes (C2) and implies that $\G$ uniquely determines $G$, regardless of the tree from which $\G$ is defined. 
However, $\G$ is tree-specific, that is, to uniquely determine $\G$ we need a pair of the measure $G$ and the finite tree $T$.
}

\begin{thm}
\label{thm: G makes X uniform}
If $X \sim G\in \cP_T$, then $\G(X) \sim {\rm Unif}((0,1]^d)$. Conversely, if $U\sim {\rm Unif}((0,1]^d)$, then $\G^{-1}(U)\sim G$.
\end{thm}
\noindent Remark: Theorem~\ref{thm: G makes X uniform} establishes (C3) and shows that if one can compute the inverse map $\G^{-1}$ then one essentially has a generative model, which allows generating samples from $G$ based on ``inverse-CDF'' sampling. More details on this will be given in Section~\ref{subsec: Forward stage-wise algorithm for learning measures}.

\subsubsection{Addition and residualization for multivariate settings}

Let $G_1,\dots,G_k$ be a collection of probability measures such that $G_l \in \cP_{T_l}$ for $l=1,2,\ldots,k$,
and let $\G_1,\dots,\G_k$ be the corresponding tree-CDFs.
As a generalization to the univariate case, next we define addition of distributions by composing their tree-CDFs. We first show that such a composition indeed pins down a unique probability measure.
\begin{lem}
    For $G_l \in \cP_{T_l}$ ($l=1,2,\ldots,k$), the mapping $F_k: \mathcal{B}(\Omega) \mapsto (0,1]$ defined as
    \begin{align}
     F_k(B) = \mu(\{\G_k \circ \cdots \circ \G_1(x):x\in B\}) \quad \text{for $B\in \mathcal{B}(\Omega)$.}
     \label{def: a summation in multi cases}
    \end{align}
    is a probability measure.
    \label{lemma: F is a probability measure}
\end{lem}
Now we can define the sum of $k$ distributions, $G_1 \oplus \cdots \oplus G_k$, as the measure $F_k$ given in Equation~\ref{def: a summation in multi cases}.
This definition of addition contains the univariate case presented earlier as a special case. We note that the addition implicitly involves the tree structures $T_1,\dots,T_k$. 
This dependency on the trees, however, is suppressed in the ``$\oplus$'' notation for simplicity without causing confusion. 

Next we turn to the notion of residuals and generalize Proposition~\ref{prop: residualizatoin for univariate cases} to multivariate distributions, which establishes Condition (C4) for tree-CDFs.
\begin{prop}
    \label{prop: we can simplify the model in multi variate cases}
    Let $G_1,\dots,G_k$ be a collection of probability measures 
such that $G_l \in \cP_{T_l}$ for $l=1,2,\ldots,k$. Then 
\begin{align*}
X \sim G_1 \oplus \cdots \oplus G_k
\quad \text{if and only if} \quad 
r^{(i)}=  \G_{i} \circ \cdots \circ \G_1(X)
    \sim
   G_{i+1}\oplus \cdots \oplus G_k
\end{align*}    
for any $i=1,2,\ldots,k-1$.
\end{prop}
\noindent Remark: This proposition implies Condition~(C4) by setting $k=2$ and $i=1$.

Moreover, the sequential update of the residuals given in Equation~\ref{eq:residuals} remains valid. 
The only difference is that now the residualization in each step depends on an implicit partition tree structure, encapsulated in the corresponding tree-CDF.

\subsection{An Unsupervised Boosting Algorithm based on Forward-stagewise (FS) Fitting}
\label{subsec: Forward stage-wise algorithm for learning measures}
Equipped with the new notions of addition and residuals, we are ready to generalize our unsupervised boosting algorithm to the multivariate setting based on forward-stagewise (FS) fitting. Suppose
we have an i.i.d.\ sample $x_1,\dots,x_n$ from an unknown distribution $F$, which we model as an additive ensemble of $K$ probability measures
\begin{align}
\label{eq:additive model}
    F=
    G_1 \oplus \cdots \oplus G_K
\end{align}
where each $G_k$ is modeled as a member in $\cP_{T_k}$ for some (unknown) $T_k$. 
We introduce an FS algorithm in which we compute the residuals step-by-step and at the $k$th step, fit $G_k$ to the current residuals. The fit at the $k$th step produces an estimate for $G_k$ along with a partition tree $T_k$, which is used to define the tree-CDF in the next step for computing the new residuals.

\noindent
\hrulefill
\begin{description}
\item[Initialization]~\\
    Set $\bm{r}^{(0)} = (x_1,\dots,x_n)$.
\item[Forward-stagewise fitting]~\\
    Repeat the following steps for $k=1,\dots,K$:
    \begin{itemize} 
    \item[1.] Fit a weak learner that produces a pair of outputs $(\G_k,T_k)$ to the residualized observations $\bm{r}^{(k-1)}$, where $T_k$ is an inferred partition tree and $\G_k$ is the tree-CDF for a measure $G_k \in \cP_{T_k}$.
    \item[2.] Update the residuals $\bm{r}^{(k)} = (r^{(k)}_1,\dots,r^{(k)}_n)$, where $r^{(k)}_i = \G_{k}(r^{(k-1)}_i)$.
    \end{itemize}
\end{description}
\hrulefill
\\

The output of the boosting algorithm in terms of the collection of pairs $(\G_k,T_k)$ for $k=1,2,\ldots,K$ contains all of the information from the data regarding the underlying distribution. (In fact the $\G_k$'s alone contain all the relevant information, but the $T_k$'s are indispensable for effectively representing and storing the $\G_k$'s.) 

Next we demonstrate two ways to extract such information. In particular, we show (i) how to compute the density function of the fitted measure $F$ at any point in the sample space analytically, and (ii) how to use the resulting generative model to draw Monte Carlo samples from the fitted measure $F$ based on ``inverse-CDF'' sampling. 

\subsubsection{Evaluating the Density Function of $F$.}

Density estimation is a common objective in learning multivariate distributions. The next proposition generalizes the additive decomposition of the log-likelihood for the univariate case and provides a recipe for evaluating the density for the fitted measure $F$ analytically based on the output of the FS algorithm.

\begin{prop}
\label{prop: simple computation of density}
For any $x \in (0,1]^d$, the density $f=dF/d\mu$ for $F$ of the additive form in Equation~\ref{eq:additive model} is given as follows
\begin{align*}
    f(x)
    &= \prod_{k=1}^{K} g_k(r^{(k-1)}) \quad \text{or} \quad \log f(x) = \sum_{k=1}^{K} \log g_k(r^{(k-1)}),
\end{align*}
where $g_k=dG_k/d\mu$ is the density of $G_k$, $r^{(k-1)}=\G_{k-1} \circ \dots \circ \G_1 (x)$, is the residual for $x$ after subtracting $G_1\oplus\cdots\oplus G_{k-1}$, and in particular $r^{(0)}=x$. In other words, the density $f(x)$ is exactly the product of the fitted density of each weak learner evaluated at the corresponding sequence of residuals. 
\end{prop}

\subsubsection{A generative model for $F$.} 
\label{subsubsec: a generative model for F}

It turns out that one can use the classical idea of inverse-CDF sampling to construct a generative model for $F$ as a result of Theorem~\ref{thm: G makes X uniform}. Specifically, we can generate samples from $F$ by first generating $U \sim {\rm Unif}((0,1]^d)$ and then compute the following transform
\begin{align}
    \F^{-1}(U) :=
    \G^{-1}_1 \circ \cdots \circ \G^{-1}_K (U), \label{eq: simulator}
\end{align}
where $\G_k^{-1}$ is the corresponding inverse for the tree-CDF $\G_k$ for $k=1,2,\ldots,K$. To implement the sampler, we next obtain the analytic form of the inverse of a tree-CDF.

Recall that in Section~\ref{subsubsec: Probability measure on a dyadic tree} we showed that a tree-CDF $\G$ for a measure $G\in \cP_T$ can be expressed as the composition of a sequence of local move mappings $\G_A:A\rightarrow A$ along each subbranch of $T$. 
The inverse of the local move mapping $\G_A$ can be expressed as $\G^{-1}_A(y) = (\G^{-1}_{A,1}(y),\dots, \G^{-1}_{A,d}(y))$ for any $y = (y_1,\dots,y_d)\in A$, where 
\begin{align*}
    \G^{-1}_{A,j}(y_j) = 
        \begin{cases}
            y_j & (j \neq j^*),\\
            \G^{'-1}_{A,j} \left(\frac{y_j - a_j}{b_j-a_j} \right) & (j = j^*),
        \end{cases}
\end{align*}
and 
\begin{align*}
     \G^{'-1}_{A,j}(z_j) = 
    \begin{cases}
        a_j + \frac{c_j - a_j}{G(A_l \mid A) }
        z_j
        & \text{ if } y_j \leq a_j + G(A_l \mid A) (b_j - a_j),\\
        c_j +
        \frac{b_j-c_j}{G(A_r \mid A)}
        \left\{
            z_j - G(A_l \mid A)
        \right\}
        & \text{ if } y_j > a_j + G(A_l \mid A) (b_j - a_j).
    \end{cases}
\end{align*}
With the inverse local move function $\G^{-1}_A$ available for all $A \in \mathcal{N}(T)$, we can obtain the explicit form for the inverse tree-CDF $\G^{-1}$ as
\begin{align*}
    \G^{-1}(y) = \G^{(R-1)} \circ \cdots \circ \G^{(1)} (y),
\end{align*}
where for $k=1,\dots,R-1,$
\begin{align*}
    \G^{(k)}(y) = \sum_{A \in \mathcal{A}^k} \G^{-1}_A(y) \1_A(y).
\end{align*}

\subsection{Decision-theoretic Considerations}
\label{subsec: Minimizing the KL divergence}
In this subsection we show that our boosting algorithm can be interpreted as fitting the additive model in Equation~\ref{eq:additive model} by sequentially reducing the Kullback-Leibler divergence.

Let $F^*$ be the true sampling distribution for the observations and $f^*=dF^*/d\mu$ its density function. Again, let $F = G_1 \oplus \cdots \oplus G_K$ be the additive model for the distribution and $f=dF/d\mu$ its density. 
We consider the entropy loss, i.e., the Kullback-Leibler (KL) divergence between $F^*$ and $F$ defined as
\begin{align}
\label{eq:entropy}
{\rm KL}(F^* || F) = \int \log \frac{f^*}{f} d F^*.
\end{align}

The next lemma states that the entropy loss in Equation~\ref{eq:entropy} can be decomposed into $K$ components (ignoring a constant) each of which only depends on $G_k$. 
\begin{lem}
    \label{lem: Decomposition of KL}
    The Kullback-Leibler divergence can be written as
    \begin{align}
        {\rm KL}(F^* || F) 
        &=
        \int \log f^* d F^*
        -
        \sum^{K}_{k=1}
        \{{\rm KL}(\tilde{F}_{k} || \mu) - {\rm KL}(\tilde{F}_{k} || G_k)\}.    
        \label{eq: decomposition of KL (population)}
    \end{align}
where $\tilde{F}_{k}$ is the true distribution of the residualized observation after subtracting $G_1,G_2,\ldots,G_{k-1}$. That is, $\tilde{F}_{k}$ is the true distribution of $r^{(k-1)}=\G_{k-1} \circ \cdots \circ \G_1(X)$, where $X \sim F^*$. 
\end{lem}
\noindent Remark: Note that because $X\sim F^*$, we have $\tilde{F}_{1}=F^{*}$, and for $k=2,\ldots, K$, $\tilde{F}_{k}$ is in the following form
    \[
        \tilde{F}_k(B) = F^*(\G^{-1}_1 \circ \cdots \circ \G^{-1}_{k-1}(B)) \quad \text{for all $B \in \mathcal{B}((0,1]^d)$}.
    \]

The first term on the right-hand side of Equation~\eqref{eq: decomposition of KL (population)} is a constant. The summand in the second term is positive as long as the measure $G_k$ is closer to $\tilde{F}_k$ than the uniform measure $\mu$ in terms of KL divergence. Hence, unless $\tilde{F}_k=\mu$, the entropy loss could be reduced by adding an additional measure $G_k$ that is closer to $\tilde{F}_k$ than $\mu$. 
In this way, fitting a measure $G_k$ to the residuals $\bm{r}^{(k)}$ in the $k$th step of our boosting algorithm can be understood as an operation to sequentially reduce the KL divergence. 
Next we turn from the above insight at the population level to the practical strategy at the finite-sample level for fitting $F$ based on $n$ i.i.d.\ observations $\{x_i\}^n_{i=1}$ from $F^*$.
First note that minimizing the divergence ${\rm KL}(F^* || F)$ is equivalent to maximizing the average log-density $\int \log f d F^*$. Thus with a finite sample, we aim to maximize the sample (average) log-density of the training data, that is,
\[
    \frac{1}{n} \sum^n_{i=1} \log f(x_i).
\]
It follows from Proposition \ref{lem: Decomposition of KL} that the log-density can also be decomposed into the sum of $K$ components, which we call ``improvements''.
\begin{lem}
\label{lem: Decomposition of KL (finite)}
The sample average log-density can be written as
\[
\frac{1}{n} \sum^n_{i=1} \log f(x_i) = \sum^K_{k=1} D^{(n)}_k(G_k),\] 
where for $k=1,2,\ldots,K$, the improvement $D^{(n)}_k(G_k)$ is
    \[
        D^{(n)}_k(G_k) = 
        \frac{1}{n} \sum^n_{i=1} \log g_k(r^{(k-1)}_i) \qquad \text{with} \qquad r^{(k-1)}_i = \G_{k-1}\circ \cdots \circ \G_1(x_i).
    \]

\end{lem}

Accordingly, the next proposition characterizes the ``optimal'' pair $(G_k,T_k)$ that maximizes $D^{(n)}_k(G_k)$. 

\begin{prop}
\label{prop: optimal solution to minimize KL (finite)}
A pair of $(G_k,T_k)$ maximizes $D^{(n)}_k(G_k)$ if and only if
   \begin{align}
        T_k \in \argmax_{T \in \T} 
        \sum_{A \in \mathcal{L}(T)}
        \tilde{F}^{(n)}_k (A)
        \log \frac{\tilde{F}^{(n)}_k(A)}{\mu(A)},
        \label{eq: condition on maximizing the average log-density (finite case)}
    \end{align}
and 
\begin{align}
G_k(A) = \tilde{F}^{(n)}_k(A) \quad \text{ for all $A\in\mathcal{L}(T_k)$},
\end{align}
where $\tilde{F}^{(n)}_k$ is the empirical measure of the residuals ${\bm r}^{(k-1)}=\{r^{(k-1)}_i\}_{i=1}^{n}$. That is,
    \[
        \tilde{F}^{(n)}_k(B)
        =
        \frac{1}{n}
        \sum^n_{i=1}
        \delta_B(r^{(k-1)}_i) \quad \text{for $B\in \mathcal{B}((0,1]^d)$.}
    \]
\end{prop}
\noindent Remark 1: The summation in Equation~\ref{eq: condition on maximizing the average log-density (finite case)} is the KL divergence between two discrete probability measures with masses given by $\{\tilde{F}^{(n)}_k(A)\}_{A \in \mathcal{L}(T)}$ and $\{\mu(A)\}_{A \in \mathcal{L}(T)}$ respectively. Equation~\ref{eq: condition on maximizing the average log-density (finite case)} implies that the ``optimal'' tree $T_k$ should allow maximal differentiation in KL divergence between the induced discretizations of $\tilde{F}^{(n)}$ and $\mu$ on its leafs.
{
This proposition offers practical guidance on how to choose a good weak learner, which will be detailed in Section~\ref{subsubsec: Choice of a Weak Learner}.
}
\\

\noindent Remark 2:
{
As suggested in Lemma~\ref{lem: Decomposition of KL (finite)}, the loss is reduced in each step as long as the improvement $D^{(n)}_k(G_k)$ is positive, and the improvement is maximized by the measure described in Proposition~\ref{prop: optimal solution to minimize KL (finite)}.
However, adopting the ``optimal'' base learner  in fitting $G_k$ as prescribed in Proposition~\ref{prop: optimal solution to minimize KL (finite)} will generally lead to over-fitting. 
As in supervised boosting \citep{hastie2009elements},
additional regularization is necessary to reduce the variance of the weak learner, and this can be achieved in analogy to supervised boosting through shrinkage toward the ``zero'', here the uniform distribution. As will be detailed in Section~\ref{subsubseq: Regularization through shrinkage}, one can still ensure the improvement to be positive when shrinkage is incorporated in an appropriate way.
}

\subsection{Group structure of tree-CDFs}
{
In the previous sections we have defined the new operation to add two distributions, and discussed the ``group-like'' structure on probability measures it induces. 
Here we make this notion more formal by showing that the collection of tree-CDFs indeed form a group. 
}

\begin{prop}
\label{prop: group structure}
{
Let $\mathcal{G}$ be a set of tree-CDFs defined as follows
\[
\mathcal{G}
=
\left\{
    \G \mid \G \text{ is a tree-CDF of } G \in \mathcal{P}_T \text{ for a finite tree }T
\right\}.
\]
Then, $\mathcal{G}$ generates a group under the composition $\circ$. 
Specifically, the identity map---which corresponds to the uniform distribution---is the identity in the group. $\mathcal{G}$ is closed under $\circ$ and each element has an inverse in $\mathcal{G}$. 
}
\end{prop}

{
This group structure is an example of the group-theoretic structure \cite{inouye2018deep} introduced to the family of density destructor transformations, though they did not show in the particular case of tree density destructors the group structure exists. 
Also note that this group is not abelian, because $\circ$ is not commutative. This is distinct from the usual group structure defined on the class of tree regressions in supervised settings, as the usual notion of addition is commutative. The forward-stepwise fitting of an additive ensemble of elements in $\mathcal{G}$ does not require the group to be abelian. 
}

\subsection{Connection to Gradient Boosting}
Many of the existing boosting algorithms can be regarded as iterative optimization of loss functions with gradient descent \citep{mason1999boosting, friedman2001greedy}.
In this subsection we discuss our new boosting from this perspective and clarify the difference from existing gradient boosting methods.

First we note that as in gradient boosting, our new algorithm can be seen as fitting a linear additive expansion.
As shown in Proposition~\ref{prop: simple computation of density}, for the ensemble measure $F = G_1 \oplus \cdots \oplus G_K$, the log-density of the ensemble measure $f = d F / d \mu$ evaluated at the observation $x_i$ ($i = 1,\dots,n$) can be decomposed as follows:
\[
\log f(x_i) = \sum_{k=1}^{K} \log g_k(r^{(k-1)}_i),
\]
where $g_k = d g_k / d\mu$ is the density of $G_k$ and $r^{(k-1)}_i = \G_{k-1} \cdots \circ \G_1 (x_i)$ is the residual. 
From this expression we can see that $\log f$ bears resemblance to a ``linear additive model'' with which we estimate the log-density function $\log f^*$ of the unknown measure $F^*$.
The fitting of this estimate to the data is evaluated with the sum of the log densities 
\[
    L(\log f)
    =
    \sum^n_{i=1}
    - \log f(x_i).
\]
Note that \cite{rosset2002boosting} also proposes an additive model for density estimation with the same objective function, but their model is a weighted sum of density functions so is different from our model. 
We also note that the input of $\log g_k$ is the residual $r^{(k-1)}_i$ instead of $x_i$ itself due to the definition of our new addition rule that involves transformation with the tree-CDFs.

Suppose that we want to update $\log f$, a current estimate of $\log f^*$, by adding a new density function $\log g$, where $g$ is the density function of the new tree measure $G$, to improve the fitting. 
The standard approach in the gradient boosting is evaluating a gradient at the current estimate and approximating its negative with the new function \citep{mason1999boosting, friedman2001greedy}. 
In our case, the gradient is constant: for all~$i$,
\[
 \left[
    \frac{\partial L}{\partial \log f (x)}
 \right]_{\log f(x) = \log f(x_i)} = - 1.
\]
Approximating its negative, 1, with the new function $\log g$ is not reasonable. 
However, this result implies that we can maximize the improvement in the loss function by maximizing the sum of the log-densities evaluated at the current residuals $\{r^{(K-1)}_i\}^n_{i=1}$,
\[
    \sum^n_{i=1}
    \log g (r^{(K-1)}_i),
\]
or equivalently their average, which is exactly what our proposed boosting algorithm iteratively does to fit the ensemble measure (see Section~\ref{subsec: Minimizing the KL divergence}).
Therefore, while our new boosting method constructs the ensemble measure in a different way from standard gradient boosting, one can still justify our algorithm as a sequential optimization of the average log densities. 

\subsection{Practical Considerations}

In this subsection we describe several practical considerations in implementing and applying the boosting algorithm. While they might first appear as technical details, we have found that they are critical in achieving competitive performance and thus worth elaborating on. Several of these considerations are drawn from similar considerations in supervised boosting.

\subsubsection{Choice of a Weak Learner}
\label{subsubsec: Choice of a Weak Learner}

Searching over all possible trees to solve Equation~\ref{eq: condition on maximizing the average log-density (finite case)} in each step of the FS algorithm is computationally prohibitive.
Nevertheless, Proposition~\ref{prop: optimal solution to minimize KL (finite)} provides hints on how to choose good weak learners that improve the KL divergence efficiently over the iterations. 
The simplest possible choice of a weak learner, as is often implemented in supervised boosting is to implement a top-down greedy tree learning algorithm that maximizes Equation~\ref{eq: condition on maximizing the average log-density (finite case)} one split at a time, as is done in fitting classification and regression trees (CART)~\citep{hastie2009elements}. 

In our numerical examples and software, we adopt a weak learner based on a simplified version of an unsupervised (Bayesian) CART model for probability distributions proposed in \cite{awaya2022hidden}. Fitting this weak learner uses a stochastic one-step look-ahead strategy to choose splitting decisions on each tree node, which generally produces closer approximation to the ``optimal'' tree splits than greedy tree algorithms. See Theorem~4.1 in \cite{awaya2022hidden} for an asymptotic justification---as the sample size grows, it produces trees that satisfy Proposition~\ref{prop: optimal solution to minimize KL (finite)} with probability increasing to~1. Additional details about the weak learner can be found in {Appendix}~\ref{supp: PT based sampling}. 

It is worth emphasizing that because we are only building ``weak'' learners that extract a small fraction of the distributional structure in each iteration, one does not need to be precisely ``optimal'' in each iteration. More importantly than being ``optimal'', the weak learner should facilitate the appropriate shrinkage to avoid overfitting, which we elaborate in the next subsection.

\subsubsection{Regularization through Scale-specific Shrinkage}
\label{subsubseq: Regularization through shrinkage}
Just as in supervised boosting, simply adopting the solution for Equation~\ref{eq: condition on maximizing the average log-density (finite case)} (either exact or approximate) as the fit for $G_k$ in each iteration will typically lead to overfitting even when the complexity of the tree $T_k$ is restricted to be small. 
In particular, the fitted density will tend to have spikes at or near the training points.  
To avoid such overfitting, it is necessary to regularize or penalize the non-smoothness in the fit for each $G_k$. 
This can be achieved through shrinkage toward ``zero'', or the uniform measure $\mu$, thereby  discounting the influence of the residuals (or its empirical measure $\tilde{F}^{(n)}_k$) on fitting $G_k$. 
In supervised boosting it is typical to introduce a learning rate $c_0\in (0,1]$ that controls how much shrinkage toward zero is applied in each iteration. In the current context, this traditional strategy would correspond to setting
\[
G_k = (1-c_0) \mu + c_0 \tilde{F}^{(n)}_{k}.
\]

We found that in practice one can further improve upon this shrinkage strategy by allowing different levels of shrinkage at different scales. The intuition is that depending on the smoothness of the underlying function, overfitting can be more (or less) likely to happen in learning local details of the distribution and thus one may benefit from enforcing a level of shrinkage that increases (or not) with the depth in the tree $T_k$. Following this intuition, we specify a scale-dependent learning rate as follows
\[
c(A)
=
c_0 \ \cdot \ (1 - \log_2 \mathrm{vol}(A))^{-\gamma},
\]
where $A$ is a node in $T_k$, $\mathrm{vol}(A)$ is a volume of $A$. Then the shrinkage toward the uniform can be specified on each node $A\in \mathcal{N}(T_k)$ in terms of the conditional probability on the children of $A$
\begin{equation}
\begin{aligned}
    G_k(A_l \mid A) &=
    (1-c(A)) \mu(A_l \mid A)
    + c(A) \tilde{F}^{(n)}_k(A_l \mid A) \ \text{ for }\  A \in \mathcal{N}(T_k), 
    \\
    G_k(\cdot \mid A) &= \mu(\cdot \mid A)\ \text{ for }\ A \in \mathcal{L}(T_k),
    \label{eq: Gk with shrinkage}
\end{aligned}
\end{equation}
where $A_l$ and $A_r$ are the children nodes of $A$ in $T_k$, $\tilde{F}^{(n)}_k(A_l \mid A) = \tilde{F}^{(n)}_k(A_l)/ \tilde{F}^{(n)}_k(A)$ if $\tilde{F}^{(n)}_k(A) > 0$ and $\tilde{F}^{(n)}_k(A_l \mid A) = \mu(A_l \mid A)$ otherwise.

The node-specific learning rate $c(A)$ controls how strongly one ``pulls'' the empirical measure $\tilde{F}^{(n)}$ toward the uniform measure $\mu$ at the corresponding scale of $A$.
It is specified with two tuning parameters $c_0 \in (0,1]$ and $\gamma \geq 0$. The parameter $c_0$ controls the global level of shrinkage, and when $\gamma > 0$ we introduce stronger shrinkage for small nodes, imposing stronger penalty on local spikes. 
When $\gamma=0$, this shrinkage reduces to the standard single learning rate specification described above. In practice, we recommend setting these tuning parameters by cross-validation.

Our next proposition shows that with shrinkage, the sample average log-density is  steadily improved in each step of the FS algorithm until the residual distribution becomes the uniform measure.
\begin{prop}
\label{prop: KL is improved even with the learning rate}
For any finite tree $T_k$, under the definition of $G_k$ given in Equation~\ref{eq: Gk with shrinkage}, the improvement satisfies $D^{(n)}_k(G_k)\geq 0$ if $c(A) \in (0,1]$ for all $A \in \mathcal{L}(T_k)$ unless $\tilde{F}^{(n)}_{k}(A)$ is indistinguishable from the uniform distribution on the tree $T_k$, that is, $\tilde{F}^{(n)}_{k}(A) = \mu(A)$ for all $A \in \mathcal{L}(T_k)$.
\end{prop}

\subsubsection{Evaluating Variable Importance}
\label{subsec: Evaluating Variable Importance}
As in supervised learning, it is often desirable to evaluate the contribution of each dimension to the approximation of the unknown measure $F^*$.
Thus we provide a way to quantify variable importance in a conceptually similar manner to what is often used in supervised boosting \citep[see][]{hastie2009elements}.
We note that \cite{ram2011density} also introduced a notion of the variable importance in density trees. While their definition is based on improvement in the $L_2$ loss, ours is based on the KL divergence, which is consistent with our earlier decision-theoretic discussion.

Specifically, because our boosting algorithm reduces the KL divergence from the unknown measure $F^*$, a natural way of quantifying the importance of a variable is adding up the decrease in the KL divergence due to splitting a tree node in the corresponding dimension.
Lemma~\ref{lem: Decomposition of KL (finite)} shows that this quantity can be expressed as the sum of the improvements $D^{(n)}_k(G_k)$.
In particular, the improvement $D^{(n)}_k(G_k)$ can be further decomposed over the splits of the tree $T_k$ as follows
\begin{align*}
    D^{(n)}_k(G_k)
    =
    \sum_{A \in \mathcal{N}(T)}
    \tilde{F}^{(n)}_k(A) 
    \left\{
        \tilde{F}^{(n)}_k(A_l \mid A) \log \frac{ G_k(A_l \mid A)}{\mu(A_l \mid A)}
        +
        \tilde{F}^{(n)}_k(A_r \mid A) \log \frac{ G_k(A_r \mid A)}{\mu(A_r \mid A)}
    \right\},
\end{align*}
where the empirical measure $\tilde{F}^{(n)}_k$ is as defined in Proposition \ref{prop: optimal solution to minimize KL (finite)}.
Note that the summation inside of the brackets can be written as
\[
    {\rm KL}(\tilde{F}^{(n)}_k(A_l \mid A) || \mu(A_l \mid A))
    -
    {\rm KL}(\tilde{F}^{(n)}_k(A_l \mid A) || G_k(A_l \mid A)),
\]
where ${\rm KL}(p || q) = p \log(p/q) + (1-p) \log[(1-p)/(1-q)]$, and it quantifies the extent to which splitting $A$ makes $G_k$ closer to the distribution of the residuals. 
Based on the decomposition, a natural definition of the total contribution of dividing in the $j$th dimension is 
\[
I_{G_k,j}=\sum_{A \in \mathcal{N}_j(T)}
    \tilde{F}^{(n)}_k(A) 
    \left\{
        \tilde{F}^{(n)}_k(A_l \mid A) \log \frac{ G_k(A_l \mid A)}{\mu(A_l \mid A)}
        +
        \tilde{F}^{(n)}_k(A_r \mid A) \log \frac{ G_k(A_r \mid A)}{\mu(A_r \mid A)}
    \right\},
\]
where $\mathcal{N}_j(T)$ represents the collection of all nodes in $T$ that are split in the $j$th dimension.
Finally we can define the importance of the $j$th variable in the additive measure $F=G_1\oplus\cdots\oplus G_K$ by summing over the variable importance across the $G_k$'s:
\begin{align*}
    I_j = \sum^K_{k=1} I_{G_k,j}.
\end{align*}

\subsubsection{Fitting the Margins and the Copula Separately and Addressing Technical Ties}
\label{subsec: two-stage algorithm}
In the density estimation literature, \cite{lu2013multivariate} suggested a two-stage strategy for estimating multivariate densities using tree-based models, which separately fits the marginal distributions and then the dependence (or copula). From our experience, this strategy can often substantially improve the fit of our unsupervised boosting algorithm. 

This two-stage strategy is easy to realize in our algorithm. In the first stage, for each of the dimensions, one can adopt weak learners that are constrained to involving tree-CDFs based on partitions along that single dimension. Computing the residuals with tree-CDFs defined on such a tree only removes the marginal distributions from the observations. ``Subtracting'' all of the marginal distributions from the original observations results in a sample of residuals the remaining distribution with uniform marginals (i.e., the corresponding copula). Then in the second stage, the single-dimension constraint on the partition trees is removed, and tree-CDFs are then fitted to the copula. The final fit is simply the sum, in terms of tree-CDF compositions, of all of the marginals and the copula. 

A related practical consideration regards tied values in the training data. (Ties in the margins occur much more frequently than ties that occur simultaneously in all margins and thus the issue is particularly relevant during the fitting of the marginal distributions in this two-stage strategy.)  When tied values occur, either from the actual data generative mechanism or due to technical reasons such as rounding, the additive tree model itself will only assume that it is due to the actual data generative mechanism and therefore there must be positive probability mass at those tied values, leading to spikes of estimated densities at those values. In practice, if the data generative mechanism is assumed to be continuous and the ties are due to technical reasons such as rounding, one can avoid this issue by a simple preprocessing step for the training data that ``smooths out'' those spikes by adding small perturbations before fitting the model. We have found a simple strategy to be effective---when there are ties in the training data at the same value $x$ and the adjacent values are $x_-$ and $x_+$ ($x_- < x < x_+$), we add uniform perturbation to the training data at $x$ on the support
$
\left(
    -(x - x_-)/2,
    (x_+ - x)/2
\right)
$.

\subsubsection{Choosing the number of trees}
\label{sec: Choosing optimal number of trees}
{
For the learning rate and the number of trees, we use the standard strategy for supervised problems \citep{friedman2001greedy} by setting the learning rate ($c_0$) small---e.g., 0.1 or 0.01---along with a large number of trees (e.g., hundreds to thousands). 
}

{
While our unsupervised boosting algorithm is generally robust with respect to overspecification in the number of trees, it is nevertheless beneficial to adopt an adaptive stopping strategy, which terminates the boosting algorithm when substantitive improvement in the fitting is no longer expected, to avoid excessive computation.
We adopt a simple strategy to achieve this. In each iteration of the algorithm, one can use a portion of the data, for example, 90\% of the residuals, to fit the next weak learner, and use the rest of the data to evaluate this tree measure by computing the average log densities.
We use this quantity to measure the improvement in the fitting. If the average improvement given by, for example, the most recent 50 trees is non-positive, we terminate the algorithm. 
We have found this adaptive stopping strategy highly effective in all of our numerical experiments---it did not affect predictive performance noticeably while substantially reducing computing time. 
}

\subsection{Expressive Power of Additive Tree Ensembles}
\label{sec: expressive power}
One interesting question is what kind of probability measures can be well approximated by the ensemble when relatively simple (e.g., shallow) tree-based weak learners are combined. This is often referred to as the ``expressive power'' of the model in the machine learning literature, or the ``support'' of the model in the statistical literature.
The expressive power for several normalizing flows have been analyzed \citep{huang2018neural, jaini2019sum, kong2020expressive}, and the large support property of linear combinations of classification trees has been established in \cite{breiman2004population}. 
We show next that a similar property holds for our unsupervised tree ensemble under the following set of conditions.
\begin{assumption}
\label{assumption to show expressive power}
For $k = 1,2,...,K$, the pair of the tree $T_k$ and the measure $G_k$ that forms a component of the tree ensemble satisfies the following conditions:
\begin{enumerate}
    \item The tree $T_k$ can be any finite tree formed by the dyadic splitting rule described in Equation~\ref{A (rectangle)} and Equation~\ref{A_l and A_r (rectangle)}. That is, it incorporates an axis-aligned splitting rule with flexible split points. 
    \item Each $T_k$ has at least $d+1$ leaf nodes, where $d$ is the dimension of the sample space. 
    \item The measure $G_k$ can be any conditionally uniform measure on $T_k$, namely, for every non-terminal node $A \in \mathcal{N}(T_k)$, the conditional probability $G_k(A_l \mid A)$ can be any value in $(0,1)$.
\end{enumerate}
\end{assumption}

Most notable in the assumption is the second condition, which is in sharp contrast to theories on single tree-based density models in the statistical literature. 
For instance, a popular tree-based density model, the \Polya tree (PT) model is shown to have the large support under the assumption that a single tree has infinite depth 
\citep{lavine1992some, ghosal2017fundamentals}.
This is not surprising in the context of additive trees, however, since Proposition 2 of \cite{breiman2004population} for supervised boosting essentially requires the same condition. With additive ensembles, we can combine small trees to express general continuous distributions, as formally stated in the next theorem.

\begin{thm}
\label{thm: universal approximation property}
Let $F^*$ be a probability measure that has a bounded density function.
Then, under Assumption \ref{assumption to show expressive power}, for any $\epsilon > 0$, there exists a tree ensemble with a finite number of tree measures, $G_1 \oplus \cdots \oplus G_K$, that approximates $F^*$ in terms of the KL divergence with this precision, i.e., 
\[
    {\rm KL}(F^* || G_1 \oplus \cdots \oplus G_K) < \epsilon.
\]
\end{thm}

\subsection{Connection to Normalizing Flows}
\label{sec: comparison with normalizing flows}
Under our definition of additive tree ensembles, the unknown distribution of the observation is modeled as the transformation of the uniform distribution that takes the form
\[
    \mathbf{T}_1 \circ \cdots \circ \mathbf{T}_K (U),\ 
    U \sim \mathrm{Unif} ((0,1]^d),
\]
where $\mathbf{T}_k = \G^{-1}_k$.
Given this expression, we can find a connection between our new boosting and the normalizing flow (NF) methods, a class of machine learning algorithms for density estimation. For comprehensive reviews of the NF, see \cite{papamakarios2019neural}, \cite{papamakarios2021normalizing}, and \cite{kobyzev2020normalizing}.
In NF methods, one approximates the observation's distribution with the transformation of known distributions such as the uniform and the Gaussian, and the transformation is represented as a composition of multiple functions.
From this viewpoint, our boosting method can be considered an NF method in which we use the inverse tree-CDFs $\G^{-1}_k$ as base transformations.

{
In NF methods, the estimated log-density is written in the form of the log-determinant of a flow transformation \citep{papamakarios2021normalizing}.
We can show that in our case, the log-determinant is identical to the sum of the log-densities that appears in our ensemble formulation.
To see this, let $\F_K$ be a composition of tree CDFs $\F_k = \G_K \circ \cdots \circ \G_1$, which uniquely corresponds to the ensemble measure written as $F = G_1 \oplus \cdots \oplus G_K$ by Theorem~\ref{thm: we can retrieve the information of a measure}.
We can rewrite the log-determinant using the density function of the tree measures $g_k = d G_k / d \mu$, as follows
\begin{align*}
\log
\mathrm{det}_{\F}(x) =
    \sum^K_{k=1}
    \log \mathrm{det}_{G_k}(r^{(k-1)}) = \sum^K_{k=1}
    \log g_k(r^{(k-1)}),
\end{align*}
where $r^{(k-1)} = \G_{k-1} \circ \cdots \circ \G_1 (x)$.
The first equation follows the chain rule, and the second equation is obtained by the fact that $\left| \mathrm{det} \G_k (x)\right| = g_k(x)$ (almost everywhere).
The last expression is the sum of the log-densities in our boosting algorithm (see Proposition~\ref{prop: simple computation of density}).
}

From an algorithmic perspective, some NF methods are similar to our boosting algorithm in that they adopt an iterative fitting approach. That is, they sequentially transform the observations to make the distribution close to the known distributions, as we sequentially residualize the observations with $\G_k$'s.
In particular, \cite{inouye2018deep} introduces an NF method called ``density destructors'', which ``substracts'' information from i.i.d. samples until they are uniform samples.  
Notably, one particular type of density destructors---the ``tree density destructors''---is defined based on subtracting a tree-based transforms, which is equivalent to our notion of tree-CDF transform. {The sequential training on the residuals is feasible due to the underlying group structure over the tree-CDF transforms. It substantially reduces the computational cost compared to methods that requires stochastic gradient descent to train.}

For most NF models, there tends to be a trade-off between the ease in evaluating the fitted density or generating samples from the fitted distribution and that in achieving large expressive power \citep{papamakarios2021normalizing}. It is worth noting that density evaluation and simulation given the fitted additive model are both straightforward to implement under the boosting algorithm because these tasks only require transforming inputs with the tree-CDF and its inverse function respectively, both of which are available in closed forms.
The computational cost of these tasks is $\mathrm{O}(R K)$ and in practice, the cost is much smaller because the node splitting is often terminated in shallow levels on much of the sample space. 
At the same time, the proposed additive tree ensemble is capable of expressing or approximating general continuous distributions as shown in Section~\ref{sec: expressive power}.

\section{Numerical Experiments}
\label{sec: Experiments}
{
In this section, we carry out numerical experiments to evaluate and demonstrate the behavior of our method.  We start with a set of simulations under several representative forms of density functions in a 48-dimensional sample space, through which we examine the impact of the tuning parameters---the learning rate, shrinkage parameters, and the number of trees---as well as the two-stage strategy on the performance of the algorithm. In addition, we compare the predictive performance of our method with a state-of-the-art single tree learner as well as the closely related tree density destructor \citep{inouye2018deep}. 
We then provide a comparative study based on several popular benchmark data sets that pitches our method against several state-of-the-art NF methods including both deep-learning based NFs and the tree density destructor.
Finally we demonstrate the computation of variable importance in both artificial datasets and the well-known MNIST handwritten digits data \citep{lecun1998gradient}.
}

{
Unless otherwise noted, we adopt the two-stage strategy and 
the adaptive stopping discussed in Section~\ref{subsec: two-stage algorithm} and in 
Section~\ref{sec: Choosing optimal number of trees}, respectively, and 
set the number of measures for the first stage (estimation of the marginal distributions) to $100$ per dimension and for the second stage (estimation of the dependence structures) to 5,000 unless otherwise statied.
The evaluation of the computational cost is done in a single-core AMD EPYC 7002 (2.50GHz) environment.
}

\subsection{Simulation Study in 48D Sample Spaces}
\begin{figure}[tb]
\centering
\begin{tabular}{c}
    \includegraphics[height=4cm]{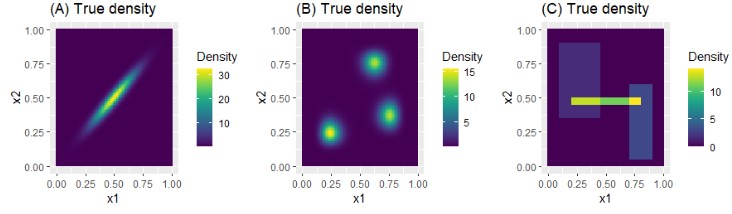}
\end{tabular}
\caption{
    The true marginal densities with respect to $X_1$ and $X_2$.
}
\label{fig: true_margins}
\end{figure}

{
We consider three 48-dimensional scenarios whose true densities are illustrated in Figure~\ref{fig: true_margins}, and set the sample size $n$ to 10,000.
(The detailed information on the data generating process is provided in Appendix \ref{sec: Details of the 48-dimensional Experiments}.)
}

\begin{figure}[tb]
\centering
\begin{tabular}{c}
    \includegraphics[height=8cm]{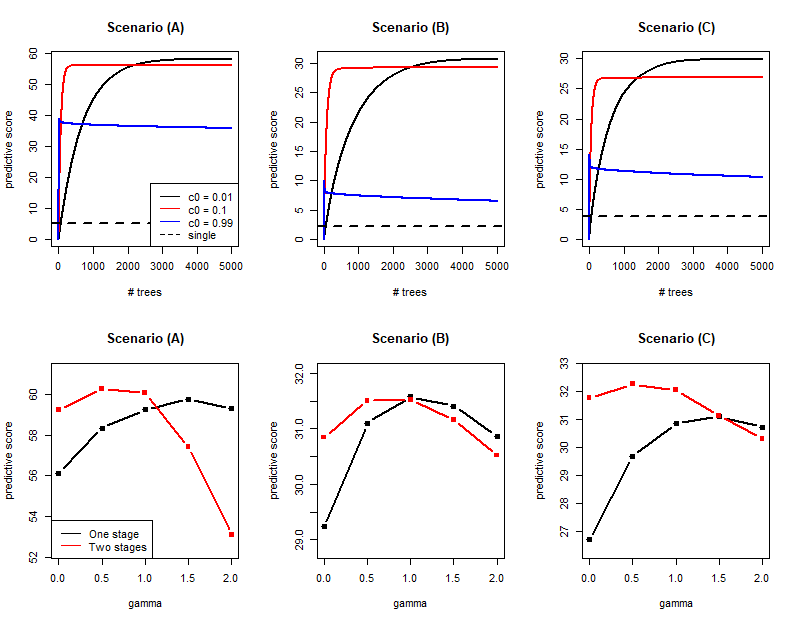}
\end{tabular}
\caption{
{
    The comparison between the single tree method and the boosting method under different tuning parameter settings.
    The first row provides the predictive performance of the single tree method and the boosting under different $c_0$ (global shrinkage parameter) and fixed $\gamma = 0.0$ (penalty for small nodes).
    The second row compares the predictive scores under different $\gamma$ values and fixed $c_0 = 0.1$.
    The shown predictive scores are averages given under 30 data sets generated with different random seeds.
}
}
\label{fig: curves (48D)}
\end{figure}

{
We evaluate the performance under two types of experimental settings. In the first experiment, we set the value of $c_0$, the global shrinkage parameter, to $0.01, 0.1$ and $0.99$ while fixing $\gamma$, the shrinkage parameter that penalizes small nodes, to 0. 
In this experiment we do not adopt the two-stage strategy and nor the adaptive stopping in order to examine the effect of changing $c_0$ and the number of trees.
The predictive performance given by the considered methods is visualized in the first row of Figure~\ref{fig: curves (48D)}. For completeness, we also report the predictive score for a state-of-the-art single tree learner (which is essentially an unregularized version of the base learner in our boosting algorithm) to illustrate the dramatic improvement in the performance through fitting an addive ensemble.
We can see that the boosting substantially outperforms the single-tree method, and that the lower learning rate ($c_0$) tends to result in better predictive scores.
Also, when $c_0$ is small (0.01), the number of trees to needed to maximize performance can be as large as 4000 to 5000.}

{
In the second experiment, we compare the performance under different $\gamma$ (the shrinkage parameter for the penalty on small nodes) and evaluate the effect of introducing the two-stage strategy.
In this experiment $\gamma$ is set to $0.0,0.5,\dots,2.0$, and $c_0$ is fixed to 0.1.
The results are reported in the second row of Figure \ref{fig: curves (48D)}. 
Overall one can see that (i) a penalty for small nodes via a positive $\gamma$ leads to improved fit as long as $\gamma$ is not too large (i.e., $\geq 1$); and (ii)  the two-stage strategy can further improve the fit.
}

\begin{figure}[tb]
\centering
\begin{tabular}{c}
    \includegraphics[height=5cm]{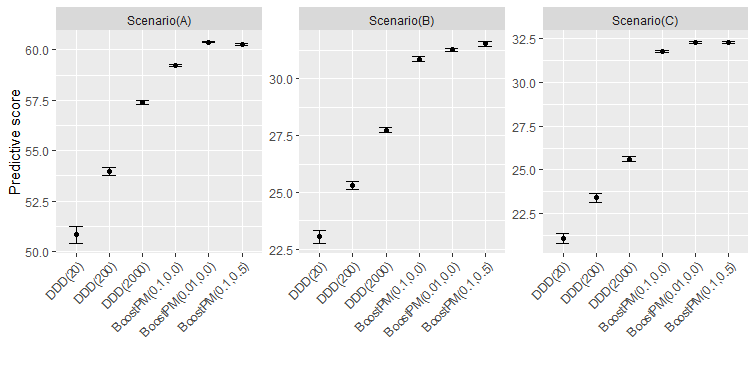}
    \\
    \includegraphics[height=5cm]{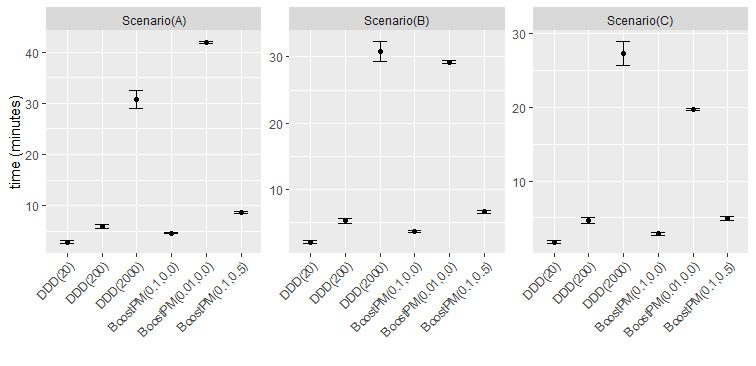}
\end{tabular}
\caption{
{
    The comparison between the boosting algorithm and the DDD (First row: predictive scores, second row: computation time) based on different 30 data sets.
    The points correspond to the means, and the intervals are made by adding/subtracting the standard deviations.
}
}
\label{fig: comparison with DDD (48D)}
\end{figure}

{
 To demonstrate the practical improvement through boosting inspired specification of our algorithm, we also provide a comparison with the deep density destructor (DDD) \citep{inouye2018deep} in terms of predictive scores and computation cost.
 For the DDD algorithm, we use the random tree destructor to fit tree measures to the residuals, which is a favorable choice in the numerical experiment provided in \cite{inouye2018deep}, and we set the shrinkage parameter ($\alpha$) to 0.2\%, 2\%, and 20\% of the sample size (20, 200, 2000) following \cite{inouye2018deep}.
 The results are provided in Figure~\ref{fig: comparison with DDD (48D)}. Our algorithm substantially outperforms the DDD  and the amount of computational time needed to achieve desirable performance is substantially less under our method.
}

\subsection{Performance Comparison with State-of-the-art Density Estimators}
\label{sec: high dim cases}

We evaluate the performance of our boosting algorithm using seven popular benchmark data sets recorded in the University of California, Irvine (UCI) machine learning repository \citep{Dua:2019}.
We preprocessed the four data sets (``POWER'', ``GAS'', ``HEPMASS'', and ``MINIBOONE'') following \cite{papamakarios2017masked} with the code provided at \url{https://github.com/gpapamak/maf} and the three data sets (``AReM'', ``CASP'', and ``BANK'') with the code provided at \url{https://zenodo.org/record/4560982#.Yh4k_OiZOCo}.
The density functions are estimated based on the training data sets, and the performance is measured by the predictive score, i.e., the average log-density evaluated at the held-out testing sets, one for each data set.

We compare our approach with three normalizing flow (NF) methods using deep neural networks to construct transforms, which represent the state-of-the-art for density estimation in machine learning: MADE \citep{germain2015made}, Real NVP \citep{dinh2016density}, and MAF \citep{papamakarios2017masked}, {and with the DDD algorithm}.
Their predictive scores are taken from \cite{papamakarios2017masked} and \citep{liu2021density}, where the detailed settings of the NF models are provided. 
{
   For the DDD algorithm, we evaluate the performance under three possible values of the shrinkage parameter ($\alpha$), 0.2\%, 2\%, and 20\% of the sample size, and show the best predictive scores.  
   For our boosting algorithm, the values of $c_0, \gamma$---the tuning parameters for regularization---are set to $(0.1, 0.0)$ and $(0.1, 0.5)$.
}

{
A comparison of the predictive scores given by the boosting and the other NF methods is provided in Table~\ref{pred scores (real data)}, where our method is labeled as ``boostPM'' (which stands for ``boosting probability measures'').
}
Our unsupervised tree boosting is overall competitive with the NF methods and even shows the best predictive performance for two data sets (``POWER'' and ``AReM'').
{Appendix~\ref{supp: tables and figures}} also provides a visual comparison of the training data sets and replicated data sets simulated from the fitted generative model, and it confirms that the distributional structures are successfully captured.
It should be noted that among the considered methods, ours is the only one that is not based on neural networks but is a combination of the tree-based learners, and therefore requires only a tiny fraction of the computational cost to train.
{
Table~\ref{table: computation time} presents the computation time on the four large data sets measured in the same single-core environment and shows our method is substantially faster to train in large $n$ settings.
Additional tables are provided in Appendix~\ref{supp: tables and figures} which show that the predictive performance of our boosting scarcely changes under different random seeds though the tree-constructing algorithm involves random splitting, and that the computation cost for simulation is much smaller compared to the cost for constructing the ensemble.}

\begin{table}[htb]
\centering
\begin{tabular}{ccccc}
 & POWER     & GAS     & HEPMASS & MINIBOONE \\ \hline
MADE & 0.40 (0.01) & 8.47 (0.02) & $\mathbf{-15.15}$ ($\mathbf{0.02}$) & $\mathbf{-12.27}$ ($\mathbf{0.47}$)     \\
RealNVP & 0.17 (0.01) & 8.33 (0.14) & -18.71 (0.02) & -13.55 (0.49)     \\
MAF & 0.30 (0.01) & $\mathbf{10.08}$ ($\mathbf{0.02}$) & $\mathbf{-17.39}$ ($\mathbf{0.02}$) & $\mathbf{-11.68}$ ($\mathbf{0.44}$)     \\
DDD & -1.91 (0.03) & -4.40 (0.31) & -28.47 (0.03) & -54.78 (1.03)     \\
BoostPM $(0.0)$ & $\mathbf{1.20}$ ($\mathbf{0.01}$) & $\mathbf{9.56}$ ($\mathbf{0.02}$) & -20.16 (0.02) & -20.38 (0.46)     \\
BoostPM $(0.5)$ & $\mathbf{1.12}$ ($\mathbf{0.01}$) & 9.42 (0.02) & -19.43 (0.02) & -17.28 (0.47)     \\
 & AReM     & CASP     & BANK &  \\ \hline
MADE & 6.00 (0.11) &  21.82(0.23) & 14.97 (0.53) &    \\
RealNVP & 9.52 (0.18) & $\mathbf{26.81}$ ($\mathbf{0.15}$) & 26.33 (0.22)   &  \\
MAF & 9.49 (0.17) & $\mathbf{27.61}$ ($\mathbf{0.13}$) & 20.09 (0.20)   &  \\
DDD &  5.73 (0.16) & 7.77 (0.19) & 9.33 (0.19)  &   \\
BoostPM$(0.0)$  & $\mathbf{12.28}$ ($\mathbf{0.16}$) & 21.84 (0.15) & $\mathbf{36.47}$ ($\mathbf{0.17}$) &      \\
BoostPM$(0.5)$ & $\mathbf{12.10}$ ($\mathbf{0.15}$) & 22.23 (0.15) & $\mathbf{36.34}$ ($\mathbf{ 0.17}$)&   
\end{tabular}
\caption{
{
The comparison of the predictive scores (MADE, RealNVP, MAF, DDD, and the proposed boosting under $\gamma=0.0, 0.5$).
The means of the estimated log-densities and the two standard deviations are provided.
The two top-performing methods for each dataset  is indicated in bold font.
}
}
\label{pred scores (real data)}
\end{table}

\begin{table}[tb]
\centering
\begin{tabular}{ccccc}
&POWER & GAS  & HEPMASS & MINIBOONE \\ \hline
$n$ & 1,659,917  &  852,174  & 315,123  &  29,556    \\
$d$ & 6 &  8  & 21 & 43   \\ \hline
MADE &   3.43 & 19.33 &  8.88 & 0.90 \\
RealNVP &   43.72 & 24.13 & 60.80 & 1.42 \\
MAF &   15.37 & 77.53 & 34.03 & 0.80\\
BoostPM(0.0) &   0.61 & 0.75 & 0.92 & 0.40\\
BoostPM(0.5) &   2.25 & 2.06 & 2.31 & 1.01
      
\end{tabular}
\caption{
 {
   The training time (hours) for the benchmark data sets, measured in minutes (standard error in parentheses) in a single-core AMD EPYC 7002 (2.50GHz) environment. 
  We show the time given under the optimal settings for MADE and RealNVP, and the time under 5 autoregressive layers for MAF. 
  Also provided are the sample size and dimensionality of the data sets.
 }
}
\label{table: computation time}
\end{table}

\subsection{Evaluating Variable Importance}
\label{sec: Evaluating variable importance (digit data)}

\begin{figure}{tb}
\centering
\begin{tabular}{c}
    \includegraphics[height=5cm]{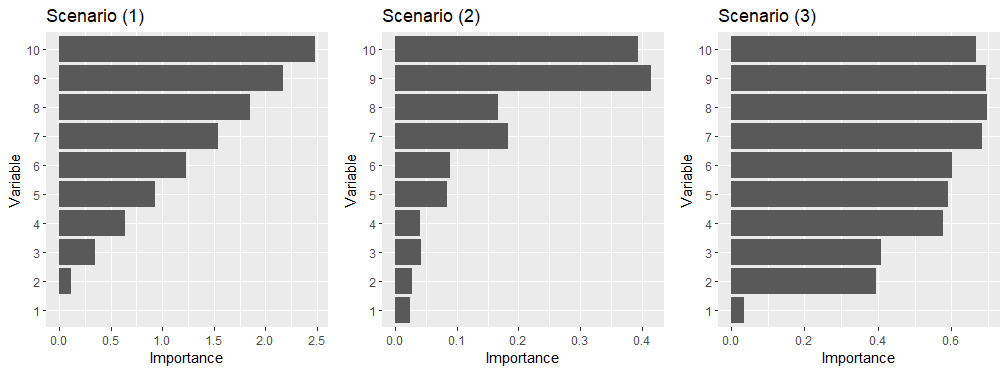}
\end{tabular}
\vspace{-0.5em}
\caption{
{
    The computed variable importance for the three simulation examples. 
}
   }
\label{fig: variable importance (simulation)}
\end{figure}

{
In this section we demonstrate the use of the variable importance measure defined in Section~\ref{subsec: Evaluating Variable Importance}.
We first carry out an experiment using artificial 10-dimensional distributions of $(X_1,\dots,X_{10})$ with $n=10,000$ using the following scenarios:
\begin{description}
    \item[Scenario(1):] $X_j \sim \mathrm{Beta}(2^{1-j}, 2^{1-j})$ for $j=1,\dots,10$.
    \item[Scenario(2):] Five pairs of random variables $(Y_{m,1}, Y_{m,2})$ ($m=1,\dots,5$), each of which follows a 2-dimensional Gaussian distribution
    \[
    \mathcal{N}
    \left(
    \left[
    \begin{array}{c}
         0  \\
         0 
    \end{array}
    \right],
    \left[
    \begin{array}{cc}
        1 & \rho  \\
        \rho & 1 
    \end{array}
    \right]
    \right),
    \]
    with $\rho = 0.1,0.3,0.5, 0.7$ and $0.9$ respectively, and
    \[
        (X_{2(m-1)+1}, X_{2m}) = 
        (\Phi(Y_{m,1}), \Phi(Y_{m,2})),
    \]
    where $\Phi(\cdot)$ is a CDF of the standard Gaussian distribution. 
    \item[Scenario(3):] Each of four groups of variables $(Y_1), (Y_2,Y_3)$, $(Y_4, Y_5, Y_6)$, and $(Y_7, Y_8, Y_9, Y_{10})$ follows the uni-variate/multi-variate Gaussian distribution in which the marginal distribution is the standard Gaussian and the correlation is 0.9, and $X_j = \Phi(Y_j)$ for $j=1,\dots,10$.
 \end{description}
The computed variable importance is displayed in Figure~\ref{fig: variable importance (simulation)}.
These results show the tendency of $X_j$ with larger $j$ obtaining larger importance, and it implies that a variable acquires higher importance when the marginal distribution is substantially different from the uniform and/or it is strongly dependent on the other variables.
}

We next compute the variable importance on the MNIST handwritten digits data \citep{lecun1998gradient}.
The gray scale images contained in this data consist of 28 $\times$ 28 $=$ 784 pixels, and each pixel takes integer values ranging from 0 (black) to 255 (white).
We obtain the data with the {\tt read\_mnist} function in the R package {\tt dslabs} \citep{dslabs} and as in \cite{papamakarios2017masked}, scale them into $[0,1]$. 
The tuning parameters $c_0$ and $\gamma$ are both set to 0.1. 

Recall that our notion of variable importance characterizes how each variable contributes to the deviation from the uniform measure in the underlying sampling distribution. For the particular application of zip-code digit recognization, a pixel is more informative about the underlying digit if it has large variation over the range of intensities. As such, the practical meaning of ``importance'' in this particular application is the opposite to the statistical importance---it is exactly those pixels with intensities spread out over large ranges (and thus more uniform) that are informative about the underlying digit. As such, we want to emphasize the difference between the ``practical importance'' and that of the ``distributional importance'' in terms of KL as we defined before.
 
\begin{figure}[htb]
\centering
\begin{tabular}{cc}
    \includegraphics[height=6.2cm]{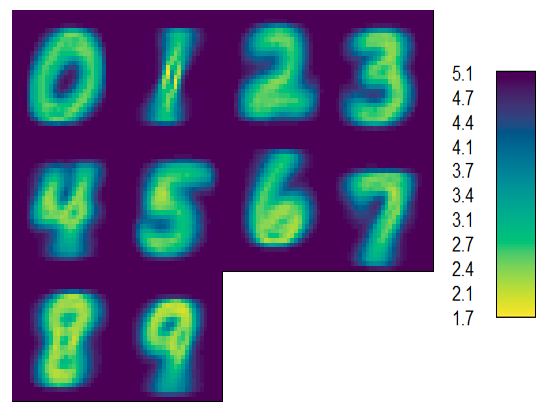}
    &
    \includegraphics[height=6.0cm]{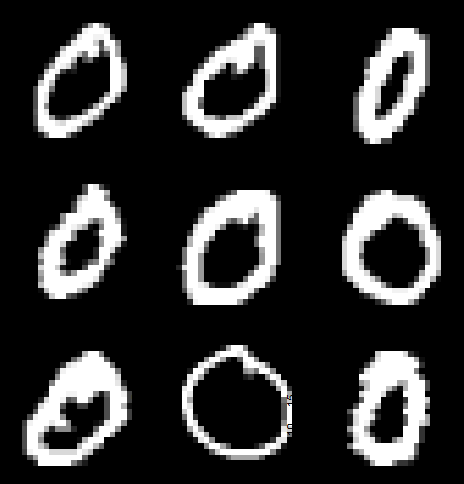}   
\end{tabular}
\caption{
    Left: The importance of pixels (variables) computed for the 10 different digits. 
    Yellow/purple colors correspond to low/high importance, which indicates large/small difference in the handwriting styles found at the pixels.
    Right: A sample of handwriting of 0.
}
\label{fig: digits}
\end{figure}

The computed ``distributional importance'' obtained for the ten different digits is visualized on the left of Figure \ref{fig: digits}, and a sample of handwriting of 0 is provided on the right.
We can see that the pixels with relatively low ``distributional importance'' and hence high practical importance lie along the outlines of the digits. 
Hence in this case the ``distributional importance'' on the left side characterizes ``the average shapes'' of the handwritten numbers.

\section{Concluding Remarks}
\label{section: conclusion}
We have proposed an unsupervised boosting method for learning multivariate probability measures by introducing new notions of addition and residuals based on tree-CDF transforms, and demonstrated how one can carry out density estimation and simulate from the fitted measure based on the output of the algorithm. Given its similarity to classical boosting for regression and classification, we expect other techniques for the boosting in such contexts, for example subsampling \citep{friedman2002stochastic}, could further improve the performance of our boosting method. Due to the limited space, we could not exploit all possible techniques in supervised boosting for improving the performance, but we expect many of them may be effective.

\section*{Acknowledgments}
LM's research is partly supported by National Science Foundation grants DMS-1749789 and DMS-2013930. NA is partly supported by a fellowship from the Nakajima Foundation. 

\bibliographystyle{plainnat} 
\bibliography{references}

\newpage

\appendix

\section{Proofs}
\label{supp: proofs}

In the following proofs, for a tree CDF $\G: (0,1]^d \mapsto (0,1]^d$ and $B \in \mathcal{B}(\Omega)$, the image is denoted by
\begin{align*}
    \G(B) = \{\G(x) : x \in B\},
\end{align*}
and the same notation rule is applied for the inverse $\G^{-1}$.
The Lebesgue measure is denoted by $\mu$.

\subsection{Proof of Proposition \ref{prop: residualizatoin for univariate cases}}
\label{supp: proof for (decomposition of CDF in 1D cases)}
By the assumption on the full support, the CDFs are invertible.

Suppose $X \sim G_1 \oplus \cdots \oplus G_k$.
For $x \in (0,1]$, we have
\begin{align*}
    P(\G_{i} \circ \cdots \circ \G_1(X) \leq x)
    &=
    P(X \leq \G_1^{-1} \circ \cdots \circ \G^{-1}_{i}(x)) \\
    &=
   \G_k \circ \cdots \circ \G_1 (\G_1^{-1} \circ \cdots \circ \G^{-1}_{i}(x)) \\
   &=
   \G_k \circ \cdots \circ \G_{i+1}(x),
\end{align*}
so $\G_{i} \circ \cdots \circ \G_1(X) \sim G_{i+1} \oplus \cdots \oplus G_k$.
The converse can be shown by transforming $P(X < x)$, where $\G_{i} \circ \cdots \circ \G_1(X)$, in the same way.

\qed

\subsection{Proof of Proposition \ref{prop: G is bijective and measurable}}
The tree CDFs are already shown to be bijective in Section \ref{subsubsec: Probability measure on a dyadic tree}, so we only show the measurability here. 

Let $\mathcal{E}$ be a set of hyper-rectangles that are written in the form of 
\[
 (a_1, b_1] \times \cdots \times (a_d, b_d]
\]
including the null set $\emptyset$.
Since $\mathcal{B}((0,1]^d)$ is the Borel $\sigma$-field, $E \in \mathcal{B}((0,1]^d)$ holds for every $E \in \mathcal{E}$.
To show the measurability of the tree CDF $\G$, which is defined by the measure $G \in \mathcal{P}_T$, it suffices to show that $\G^{-1}(E) \in \mathcal{B}((0,1]^d)$ for every $E \in \mathcal{E}$ since $\mathcal{E}$ generates $\mathcal{B}((0,1]^d)$.

By the definition of $\G$, the image $\G(A)$ ($A \in \mathcal{L}(T)$) is also a hyper-rectangle included in $\mathcal{E}$, and their collection 
$
    \{\G(A): A \in \mathcal{L}(T)\}
$
forms a partition of $(0,1]^d$.
Hence $E$ is written as a union of disjoint sets:
\[
    E = \bigcup_{A \in \mathcal{L}(T)}
    (E \cap \G(A)),
\]
where each $E \cap \G(A)$ is a hyper-rectangle that belongs to $\mathcal{E}$ and a subset of $\G(A)$.
Hence $\G^{-1}(E \cap \G(A))$ also belongs to $\mathcal{E} \subset \mathcal{B}((0,1]^d)$.
Therefore, their finite union 
\[
    \G^{-1}(E) = \bigcup_{A \in \mathcal{L}(T)}
    \G^{-1}(E \cap \G(A))
\]
is also an element of $\mathcal{B}((0,1]^d)$. 
\qed

\subsection{Proof of Theorem \ref{thm: G makes X uniform}}
To prove the first assertion, define $\G^{[r]}$ for $r = 1,\dots,R-1$ as 
\[
    \G^{[r]}(x)
    =
    \sum_{A \in \A^r}
    \G_A(x) \1_A(x),
\].
Then the first assertion is equivalent to that if $X \sim G$, then
\begin{align}
    \G^{[1]} \circ \dots \circ \G^{[R-1]} (X) \sim Unif((0,1]^d),
    \label{eq: conclusion of the first theorem}
\end{align}
which we prove here.

In the proof, we let $X^{[R]} = X$ and for $r=1,\dots,R-1$
\begin{align*}
    X^{[r]}
    &=
    \G^{[r]}   
    \circ
    \cdots
    \circ
    \G^{[R-1]}
    (X).
\end{align*}
Let $G^{[r]}$ denote a probability measure for $X^{[r]}$.
With these notations, we prove \eqref{eq: conclusion of the first theorem} by induction:
We show that, if for $A \in \mathcal{N}(T)$, and 
\begin{align}
    &G^{[r+1]}(A_l) = G(A_l),\  
    G^{[r+1]}(A_r) = G(A_r), \notag
    \\
    &G^{[r+1]}(\cdot \mid A_l) = \mu(\cdot \mid A_l),\ 
    G^{[r+1]}(\cdot \mid A_r) = \mu(\cdot \mid A_r),
    \label{eq: condition in induction}
\end{align}
then
\begin{align}
   G^{[r]}(A) = G(A),\ G^{[r]}(\cdot \mid A) = \mu(\cdot \mid A).
     \label{eq: condition in induction 2}
\end{align}
The conditions in Equation~\ref{eq: condition in induction} holds if $r = R-1$ because $G^{[R]} = G$ and $G \in \cP_T$, and the statement in Equation~\ref{eq: condition in induction 2} being true for $r=1$ implies that 
$\G^{[1]}   
    \circ
    \cdots
    \circ
\G^{[R]}(\cdot) = \mu(\cdot),
$
which is equivalent to Equation~\ref{eq: conclusion of the first theorem}.

Assume Equation~\ref{eq: condition in induction} holds for some $r$.
By the definition, $\G^{[r]}$ is bijective, and $\G^{[r],-1}(A) = A$ for every $A \in \mathcal{A}^r$. Then for $X \sim G^{[r]}$ and $A \in \mathcal{A}^r$, we have,
\begin{align*}
    G^{[r]} (A) 
    &= P(X^{[r]} \in A) \\ 
    &=
    P(\G^{[r],-1}(X^{[r]}) \in \G^{[r],-1}(A)) 
    =
    P(X^{[r+1]} \in A)  \\
    &=
    G^{[r+1]}(A_l) + G^{[r+1]}(A_r) = G(A).
\end{align*}
Hence the first equation in Equation~\ref{eq: condition in induction 2} holds. 
To prove the second equation, let $X^{[r]} = (X^{[r]}_1,\dots,X^{[r]}_d)$.
Then, we show that for $z_j \in (a_j, b_j]$,
\begin{align}
    &P(X^{[r]}_1 \in (a_1, z_1], \dots, X^{[r]}_d \in (a_d, z_d] \mid X^{[r]} \in A) =
    \prod^d_{j=1} \frac{z_j - a_j}{b_j - a_j}
    \label{is uniform}
\end{align}
holds.
The probability in the left hand side can be written as follows:
\begin{align}
    &P(X^{[r]}_1 \in (a_1, z_1], \dots, X^{[r]}_d \in (a_d, z_d] \mid X^{[r]} \in A) \notag \\
    &= P(X^{[r]}_1 \in (a_1, z_1], \dots, X^{[r]}_d \in (a_d, z_d] \mid X^{[r+1]} \in A) \notag \\
    &=
    \frac{P(X^{[r+1]} \in A_l)}{P(X^{[r+1]} \in A)}
    P(X^{[r]}_1 \in (a_1, z_1], \dots, X^{[r]}_d \in (a_d, z_d] \mid X^{[r+1]} \in A_l) \notag \\
    &\ +
    \frac{P(X^{[r+1]} \in A_r)}{P(X^{[r+1]} \in A)}
    P(X^{[r]}_1 \in (a_1, z_1], \dots, X^{[r]}_d \in (a_d, z_d] \mid X^{[r+1]} \in A_r) \notag \\
    &=
    G(A_l \mid A)
    P(X^{[r]}_1 \in (a_1, z_1], \dots, X^{[r]}_d \in (a_d, z_d] \mid X^{[r+1]} \in A_l) \notag \\
    &\ +
    G(A_r \mid A)
    P(X^{[r]}_1 \in (a_1, z_1], \dots, X^{[r]}_d \in (a_d, z_d] \mid X^{[r+1]} \in A_r).
    \label{second term}
\end{align}
Let $A$ be divided in the $j^*$th dimension.
By the definition of $\G_A$, for $j \neq j^*$, $X^{[r]}_j \in (a_j, z_j] \Longleftrightarrow X^{[r+1]}_j \in (a_j, z_j]$. 
For $j^*$, because $\G_{A,j^*}(\cdot)$ is strictly increasing, 
\begin{align*}
    X^{[r]}_{j^*} \in (a_{j^*}, z_{j^*}] 
    &\Longleftrightarrow
    \G^{-1}_{A,j^*}
    \left(
        X^{[r]}_{j^*}
    \right)
    \in
    \left(
        \G^{-1}_{A,j^*}(a_{j^*}),
        \G^{-1}_{A,j^*}
        \left(
            z_{j^*}
        \right)
    \right] \\
    &\Longleftrightarrow  
    X^{[r+1]}_{j^*}
    \in
    \left(
        a_{j^*},
        y_{j^*}
    \right],
\end{align*}
where $    y_{j^*} =
    \G^{-1}_{A,j^*}
    (
        z_{j^*} 
    )$.
The expression of $y_{j^*}$ changes depending on whether $y_{j^*} \leq c_{j^*}$ or not, where $c_{j^*}$ is a partition point at which $A$ is divided.
We first assume that $y_{j^*} \leq c_{j^*}$.
In this case, the second term in Equation~\ref{second term} is 0 because $X^{[r+1]}_{j^*}
    \in
    \left(
        a_{j^*},
        y_{j^*}
    \right]$ does not happen if $X^{[r+1]} \in A_r$. 
Also, by the definition of $\G_{A,j^*}$, 
\begin{align*}
\frac{z_{j^*} - a_{j^*}}{y_{j^*} - a_{j^*}}
&=\frac{G(A_l \mid A))}{\mu(A_l \mid A)}
=
G(A_l \mid A))
\frac{b_{j^*}-a_{j^*}}{c_{j^*}-a_{j^*}}
\\
\Longleftrightarrow
\frac{y_{j^*} - a_{j^*}}{c_{j^*} - a_{j^*}}
&=
\frac{1}{G(A_l \mid A))} \frac{z_{j^*} - a_{j^*}}{b_{j^*} - a_{j^*}}.
\end{align*}
Therefore, it follows that
\begin{align*}
    &P(X^{[r]}_1 \in (a_1, z_1], \dots, X^{[r]}_d \in (a_d, z_d] \mid X^{[r]} \in A) \\
    &=
    G(A_r \mid A)
    \left\{
        \prod_{j\neq j^*} \frac{z_j - a_j}{b_j - a_j}  
    \right\}
    \frac{y_{j^*} - a_{j^*}}{c_{j^*} - a_{j^*}} \\
    &=
    G(A_r \mid A)
    \left\{
        \prod_{j\neq j^*} \frac{z_j - a_j}{b_j - a_j}  
    \right\}
    \frac{1}{G(A_l \mid A))} \frac{z_{j^*} - a_{j^*}}{b_{j^*} - a_{j^*}} \\
    &=
    \prod^d_{j = 1} \frac{z_j - a_j}{b_j - a_j}. 
\end{align*}
We can prove (\ref{is uniform}) for the case of $y_{j^*} > c_{j^*}$ in the same way.
\\

To prove the second result, let $U \sim \mathrm{Unif}((0,1]^d)$.
The multi-scale CDF $\G$ is bijective (Proposition \ref{prop: G is bijective and measurable}), so we obtain for $B \in \mathcal{B}(\Omega)$
\begin{align*}
    P(\G^{-1}(U) \in B)
    &=
   P(U \in \G(B))\\
   &=
   \mu(\G(B)) \\
   &=
   G(B),
\end{align*}
where the last line follows Theorem \ref{thm: we can retrieve the information of a measure}.
(Note that the proof of  Theorem \ref{thm: we can retrieve the information of a measure} only uses the first result of Theorem \ref{thm: G makes X uniform}).
Therefore $\G^{-1}(U) \sim G$.
\qed 

\subsection{Proof of Theorem \ref{thm: we can retrieve the information of a measure}}
Let $X \sim G$.
By the first result of Theorem \ref{thm: G makes X uniform}, we obtain
\begin{align*}
    G(B) &= P(X \in B) \\
    &= P(\G(X) \in \{\G(x) : x \in B\})\\
    &= \mu(\{\G(x) : x \in B\}). \qed
\end{align*}

\subsection{Proof of Lemma \ref{lemma: F is a probability measure}}
We only need to check the countable additivity.
By Proposition \ref{prop: G is bijective and measurable}, $\G_k \circ \cdots \circ \G_1$ is bijective. 
Hence, for disjoint sets $A_l \in \mathcal{B}(\Omega)\ (l \in \mathbb{N})$, it follows that
    \[
        \G_k \circ \cdots \circ \G_1\left(
            \bigcup_{l} A_l
        \right) 
        =
        \bigcup_{l}
        \G_k \circ \cdots \circ \G_1 \left(
             A_l
        \right).
    \]
    Because $\{\G_k \circ \cdots \circ \G_1 \left(
             A_l
        \right)\}_{l=1,2,\dots}$ are disjoint, this result implies that
    \begin{align*}
         F_k\left(
            \bigcup_{l} A_l
        \right)
         &=
        \mu\left(
        \bigcup_{l}
        \G_k \circ \cdots \circ \G_1 \left(
             A_l
        \right)
        \right) \\
        &=
        \sum_l\mu(\G_k \circ \cdots \circ \G_1\left(
             A_l
        \right))\\
        &=
        \sum_l F \left(
             A_l
        \right). \qed       
    \end{align*}

\subsection{Proof of Proposition \ref{prop: we can simplify the model in multi variate cases}}
First we suppose $X \sim G_1 \oplus \cdots \oplus G_k$.
By Proposition \ref{prop: G is bijective and measurable}, $\G_1,\dots,\G_k$ are all bijective, so for $i=1,\dots,k$, the composition 
\[\F_i = \G_i \circ \cdots \circ \G_1\]
is also bijective.
For $B \in \mathcal{B}(\Omega)$, we obtain
\begin{align*}
    P(\F_i(X) \in B)
    &=
    P(X \in \F^{-1}_i(B)) \\
    &=
    F_k((\F^{-1}_i(B)))\\
    &=
    \mu(\F_k(\F^{-1}_i(B))) \\ 
    &=\mu(
    \G_{k} \circ \cdots \circ \G_{i+1}
    (B)). 
\end{align*}
Hence $\F_i(X) \sim G_{i+1} \oplus \cdots \oplus G_k$.
Showing the converse is now straightforward. 
\qed

\subsection{Proof of Proposition \ref{prop: simple computation of density}}

We first show the following lemma, which implies that the conditional distributions $F_{k-1}$ and $F_k$ ($k=2,\dots,K$) are the same on subsets in a partition defined by $T_k$ and $\F_{k-1} = \G_{k-1} \circ \cdots \circ \G_1$.

\begin{lem}
\label{lem: conditional distributions of Fk-1 and Fk}
Let $\mathcal{L}(T_k)$ be a set of the terminal nodes in $T_k$.
Also, we let $A' \in \mathcal{L}(T_k)$ and $A = \F^{-1}_{k-1}(A')$. 
Then for any $B \subset A$, $F_{k-1}(B \mid A) = F_k(B \mid A)$.
\end{lem}
\noindent
(Proof)
For $B \subset A$, by Theorem \ref{thm: we can retrieve the information of a measure}, we have
\begin{align}
    F_k(B)
    &=
    \mu(\F_k (A)) \notag \\
    &=
    \mu(\G_k(\F_{k-1}(B))) \notag \\
    &=
    G_k(\F_{k-1}(B)).
    \label{Transformation for B in A}
\end{align}
Since $B \subset A$, $\F_{k-1}(B) \subset \F_{k-1}(A) = A'$.
Hence $F_k(B)$ is further rewritten as follows
\begin{align*}
    F_k(B)
    &=
    G_k(A' \cap \F_{k-1}(B)) \\
    &=
    G_k(A') G_k(\F_{k-1}(B) \mid A') \\
    &=
    G_k(A') \mu(\F_{k-1}(B) \mid A') \\
    &=
    G_k(A') \frac{\mu(\F_{k-1}(B))}{\mu(A')}\\
    &=
    G_k(A') \frac{F_{k-1} (B)}{ \mu(A')}.
\end{align*}
By the definition of $A$, $F_{k-1}(A) = \mu(\F_{k-1}(A)) = \mu(A')$, and by replacing $B$ with $A$ in (\ref{Transformation for B in A}), we have
\begin{align*}
    F_k(A) = G_k(\F_{k-1}(A)) = G_k(A').
\end{align*}
Therefore, we obtain
\begin{align*}
    F_k(B \mid A)
    &=
    \frac{F_k(B)}{F_k(A)}\\
    &=
    G_k(A') \frac{F_{k-1} (B)}{ \mu(A')}
    \frac{1}{F_k(A)} \\
    &=
    \frac{F_{k-1}(B)}{F_{k-1}(A)}\\
    &=
    F_{k-1}(B \mid A).
\end{align*}
\qed

\noindent
(Proof of Proposition \ref{prop: simple computation of density})
Let $\mathcal{L}_k = \{ \F^{-1}_{k-1}(A'):A' \in \mathcal{L}(T_k)\}$.
By Lemma \ref{lem: conditional distributions of Fk-1 and Fk}, the conditional distributions $F_{k-1}(\cdot \mid A)$ and $F_k(\cdot \mid A)$ are the same for $A \in \mathcal{L}_k$.
Hence the density functions of $F_{k-1}$ and $F_k$ denoted by $f_{k-1}$ and $f_k$ are expressed as
\begin{align*}
    f_{k-1} (x)
    &=
    \sum_{A \in \mathcal{L}_k}
    F_{k-1}(A) f_{k-1} (x \mid A) \1_A(x), \\
    f_k (x)
    &=
    \sum_{A \in \mathcal{L}_k}
    F_k(A) f_{k-1} (x \mid A) \1_A(x),
\end{align*}
where $\1$ is the indicator function.
Fix $x \in (0,1]^d$ and let $A_k \in \mathcal{L}_k$ such that $x \in A_k$ and $A'_k = \F_{k-1}(A_k)$.
By Theorem \ref{thm: we can retrieve the information of a measure}, we have
\begin{align*}
    F_{k-1}(A_k)
    &=
    \mu(\F_{k-1}(A_k)) = \mu(A'_k),\\
    F_k(A_k)
    &=
    \mu(\F_k(A_k))\\
    &=
    \mu(\G_k \circ \F_{k-1}(A_k)) \\
    &=
    G_k(\F_{k-1}(A_k)) \\
    &= G_k(A'_k).
\end{align*}
Hence, we have 
\begin{align}
    \frac{f_k(x)}{f_{k-1}(x)}
    &=
    \frac{F_k(A_k)}{F_{k-1}(A_k)}
    = \frac{G_k(A'_k)}{\mu(A'_k)}.
    \label{density ratio}
\end{align}
Since $x\in A_k$, 
\begin{align*}
    \F_{k-1}(x) 
    =
    \G_{k-1} \circ \cdots \circ \G_1 (x)
    \in \F_{k-1}(A_k) = A'_k.
\end{align*}
Thus the density ratio in (\ref{density ratio}) is rewritten as
\begin{align*}
    \frac{f_k(x)}{f_{k-1}(x)}
    &=
    G_k(A'_k)
    \mu(\G_{k-1} \circ \cdots \circ \G_1 (x) \mid A'_k) \\
    &=
    G_k(A'_k)
    g_k(\G_{k-1} \circ \cdots \circ \G_1 (x) \mid A'_k) \\
    &=
    g_k(\G_{k-1} \circ \cdots \circ \G_1 (x)),
\end{align*}
where the second equation follows that $A'_k \in \mathcal{L}(T_k)$.
Because the discussion above holds for $k=2,\dots,K$, we obtain the following expression
\begin{align*}
    f_K(x)
    &=
    f_1(x)
    \prod^K_{k=2}
    \frac{f_k(x)}{f_{k-1}(x)} \\
    &=
    f_1(x)
    \prod^K_{k=2}
    g_k(\G_{k-1} \circ \cdots \circ \G_1 (x)).  
\end{align*}
\qed

\subsection{Proof of Proposition \ref{prop: group structure}}
{
The associativity clearly holds, and showing the existence of the identity element is also straightforward because the identity transformation, which is a tree-CDF of the uniform distribution, is included in $\mathcal{G}$. \\
Proving the existence of an inverse element for every element in $\mathcal{G}$ is done by showing that an inverse function of a local-move function, which is an essential component of a tree-CDF, is also a local-move function. 
To this end we use the local-move function $\G_A: A \mapsto A$ defined in Section~\ref{subsubsec: Probability measure on a dyadic tree} and the same notations. \\
Let $\tilde{c}_{j^*} = a_{j^*} + G(A_l \mid A) (b_{j^*} - a_{j^*})$ and $\tilde{A}_l$ and $\tilde{A}_r$ be a pair of nodes that we obtain by dividing $A$ at $\tilde{c}_{j^*}$ in the $j^*$th dimension.
Then we define a function $\tilde{\G}_{A}: A \mapsto A$ such that for any $x\in A$, $\tilde{\G}_A(x) = (\tilde{\G}_{A,1}(x),\dots,\tilde{\G}_{A,d}(x))$ where $\tilde{\G}_{A,j}(x)=x$ for all $j\neq j^*$, and
\[
\frac{\tilde{\G}_{A,j^*}(x)-a_{j^*}}{x_{j^*}-a_{j^*}} = \frac{\mu(A_l|A)}{\mu(\tilde{A}_l|A)} \quad \text{for $x\in \tilde{A}_l$}
\quad
\text{and}
\quad 
\frac{b_{j^*}-\tilde{\G}_{A,j^*}(x)}{b_{j^*}-x_{j^*}} = \frac{\mu(A_r|A)}{\mu(\tilde{A}_r|A)} \quad \text{for $x\in \tilde{A}_r$}.
\]
From this definition $\tilde{\G}_A$ works as a local-move function for the conditional distribution that assigns the probability $\mu(A_l|A)$ to $\tilde{A}_l$.
It is straightforward to see that this transformation $\tilde{\G}_A$ is identical to $\G_A^{-1}$ (see the expression of the inverse function provided in Section~\ref{subsubsec: a generative model for F}). 
}
\qed
\\

\subsection{Proof of Lemma \ref{lem: Decomposition of KL}}
By the definition of the KL divergence, we have
\begin{align*}
            {\rm KL}(F^* || F) 
         &=
         \int \log f^* d F^*
         -
         \int \log f d F^*.
\end{align*}
By Proposition \ref{prop: simple computation of density}, the second term $\int \log f F^*$ is decomposed as
\begin{align*}
    &\int \log f dF^*
    =
    \sum^{K}_{k=1} \int \log g_k(\G_{k-1} \circ \dots \circ \G_1 (x)) d F^*(x).
\end{align*}
By the change-of-variable formula (e.g., Theorem 3.6.1 in \cite{bogachev2007measure}), the right hand side can be written in a form of integration with respect to $\tilde{F}_k$,
\begin{align*}
    \int \log g_k(\G_{k-1} \circ \dots \circ \G_1 (x)) d F^*(x)
    =
     \int \log g_k(x) d \tilde{F}_k(x).
\end{align*}
Since the measure $F^*$ is absolutely continuous with respect to the Lebesgue measure $\mu$, by the definition of $\tilde{F}_k$, that is,
    \[
        \tilde{F}_k(B) = F^*(\G^{-1}_1 \circ \cdots \circ \G^{-1}_{k-1}(B)) \quad \text{for all $B \in \mathcal{B}((0,1]^d)$},
    \]
$\tilde{F}_k$ is also absolutely continuous with respect to $\mu$.
Hence $\tilde{F}_k$ admits the density function denoted by $\tilde{f}_k(x)$, and the right hand side is further rewritten as follows
\begin{align*}
    \int \log g_k(x) d \tilde{F}_k(x) 
    &=
    \int \log \frac{\tilde{f}_k(x)}{\mu(x)} d \tilde{F}_k(x)
    -
    \int \log  \frac{\tilde{f}_k(x)}{g_k(x)} d \tilde{F}_k(x)\\
    &=
    {\rm KL}(\tilde{F}_{k} || \mu) - {\rm KL}(\tilde{F}_{k} || G_k). \qed
\end{align*}

\subsection{Proof of Proposition \ref{prop: optimal solution to minimize KL (finite)}}
\label{subsec: Proof of Proposition (optimal solution to minimize KL (finite))}
Since $T_k$ is a finite tree, the log of $g_k$, which is piece-wise constant on $T_K$, is  written as
\[
    \log g_k(x)
    = 
    \sum_{A \in \mathcal{L}(T_k)}
    \log \frac{G_k(A)}{\mu(A)} \1_A(x)
    \quad 
    \text{for } x \in (0,1]^d.
\]
Hence the improvement $D^{(n)}_k(G_k)$ is rewritten as follows:
\begin{align*}
    D^{(n)}_k(G_k)
    &=
    \sum_{A \in \mathcal{L}(T_k)}
    \tilde{F}^{(n)}_k (A)
    \log \frac{G_k(A)}{\mu(A)} \\
    &=
    \sum_{A \in \mathcal{L}(T_k)}
    \tilde{F}^{(n)}_k (A)
    \log \frac{\tilde{F}^{(n)}_k(A)}{\mu(A)}
    -
    \sum_{A \in \mathcal{L}(T_k)}
    \tilde{F}^{(n)}_k (A)
    \log \frac{\tilde{F}^{(n)}_k(A)}{G_k(A)}. 
\end{align*}
Because the second term in the bottom line takes a form of KL divergence defined for the two discrete distributions, it is minimized if $G_k(A) = \tilde{F}^{(n)}_k (A)$ for all $A \in \mathcal{L}(T_k)$.
Under this $G_k$, since the second term is 0, the improvement is maximized if $T_k$ satisfies the condition provided in Proposition  \ref{prop: optimal solution to minimize KL (finite)}.
\qed

\subsection{Proof of Proposition \ref{prop: KL is improved even with the learning rate}}
In this proof, we suppose the learning rate $c(A)$ is independent to a node $A$ for simplicity.
For every leaf node $A \in \mathcal{L}(T_k)$, there is a sequence of nodes $\{B_{A, r}\}^R_{r=1}$ such that $B_{A, r}$ belongs to the $r$th level of $T_k$, and 
\[
    (0,1]^d = B_{A, 1} \supset  B_{A, 2} \supset \cdots \supset B_{A, r} = A.
\]
With such sequences, based on the discussion in {Appendix \ref{subsec: Proof of Proposition (optimal solution to minimize KL (finite))} }, the improvement $D^{(n)}_k(G_k)$ is decomposed as
\begin{align*}
    D^{(n)}_k(G_k)
    &=
    \sum_{A \in \mathcal{L}(T_k)} \tilde{F}^{(n)}_k (A) \log  \frac{G_k(A)}{\mu(A)} \\
    &=
    \sum_{A \in \mathcal{L}(T_k)} \tilde{F}^{(n)}_k (A) 
    \left[
    \log  \frac{G_k(B_{A, 2} | B_{A, 1})}{\mu(B_{A, 2} | B_{A, 1})}
    + \cdots +
    \log  \frac{G_k(B_{A, R} | B_{A, R-1})}{\mu(B_{A, R} | B_{A, R-1})}
    \right]
    \\
    &=
    \sum_{A \in \mathcal{N}(T_k)}
    \left[
        \tilde{F}^{(n)}_k (A_l) \log \frac{G_k(A_l \mid A)}{\mu(A_l \mid A)}
        +
        \tilde{F}^{(n)}_k (A_r) \log \frac{G_k(A_r \mid A)}{\mu(A_r \mid A)}
    \right].
\end{align*}
For the bottom line the summand is 0 if $\tilde{F}^{(n)}_k(A) = 0$.
Otherwise, the conditional probabilities $\tilde{F}^{(n)}_k(A_l \mid A)$ and $\tilde{F}^{(n)}_k(A_r \mid A)$ are defined.
In such a case, by the definition of $G_k(A_l \mid A)$,
\begin{align*}
    \log \frac{G_k(A_l \mid A)}{\mu(A_l \mid A)}
    &=
    \log
    \left[
        \frac{(1-c) \mu(A_l \mid A)
        + c \tilde{F}^{(n)}_k(A_l \mid A)}{\mu(A_l \mid A)}
    \right]\\
    &\geq
    (1-c) \log 1 + c \log \frac{\tilde{F}^{(n)}_k(A_l \mid A)}{\mu(A_l \mid A)}\\
    &=c \log \frac{\tilde{F}^{(n)}_k(A_l \mid A)}{\mu(A_l \mid A)},
\end{align*}
where the second line follows the Jensen's inequality.
The same result holds for $A_r$.
Hence,
\begin{align*}
    &  \tilde{F}^{(n)}_k (A_l) \log \frac{G_k(A_l \mid A)}{\mu(A_l \mid A)}
        +
        \tilde{F}^{(n)}_k (A_r) \log \frac{G_k(A_r \mid A)}{\mu(A_r \mid A)}
    \\
    &
    \geq c \tilde{F}^{(n)}_k (A) 
    \left[
        \tilde{F}^{(n)}_k(A_l \mid A)
        \log \frac{\tilde{F}^{(n)}_k(A_l \mid A)}{\mu(A_l \mid A)}
        +
        \tilde{F}^{(n)}_k(A_r \mid A)
        \log \frac{\tilde{F}^{(n)}_k(A_r \mid A)}{\mu(A_r \mid A)}
    \right],
\end{align*}
where the sum inside of the brackets is the KL divergence for the two Bernoulli distributions and thus non-negative. Therefore, the improvement $D^{(n)}_k(G_k)$ is non-negative.
Additionally, the last inequality is strict if and only if $\tilde{F}^{(n)}_k(A_l \mid A) = \mu(A_r \mid A)$ and so  $D^{(n)}_k(G_k)$ is positive.
\qed

\section{Expressive Power of the Tree Ensemble}
\label{sec: Expressive power of the tree ensemble}
In this section, we provide theoretical results on the expressive power of the tree ensemble with the final goal of proving Theorem \ref{thm: universal approximation property}.

\subsection{Preparations}

We introduce the following notations:
\begin{enumerate}
\item Let $\T^{L}$ be a collection of dyadic trees with axis-aligned boundaries with at most $L$ maximum resolution.
When $L = d$, $\T^{d}$ is a set of trees that can be formed under Assumption~\ref{assumption to show expressive power}.
We note that as implied in the following proofs, $\T^{L}$ can a set of trees that have at least one node reach the $L$th while the other leaf nodes belong to the shallower levels.

\item For a tree $T \in \T^L$, a set $\mathcal{P}_T$ denotes a collection of probability measures conditionally uniform on $T$ such that 
\begin{align}
    G(\cdot \mid A) &= \mu(\cdot \mid A) \text{ and } G(A) > 0
\end{align}
for every terminal node $A \in T$.
A collection of such tree measures are denoted by $\cG^L_0$, that is,
\[
    \cG^L_0 = \{
        G :
        G \in \mathcal{P}_T \text{ for some } T \in \T^L
    \}.
\]
For a measure $G \in \cG^L_0$ defined on a tree $T \in \T^L$, we can define a tree-CDF as in Section \ref{sec: method}, which is denoted by $\G$.
We define a set $\bG^L_0$ as a collection of such tree CDFs, namely, 
\[
    \bG^L_0 = \{
        \G:
        \G \text{ is a tree CDF of } G \in \cG^L_0
    \}.
\]
\item Let $\bG^L$ denote a set of finite composition of tree CDFs, that is,
\[
    \bG^L
    =
    \{
        \G_K \circ \cdots \circ \G_1 :
        K \in \mathbb{N}
        \text{ and for } k =1,\dots,K,\ 
        \G_k \in \bG^L_0
    \},
\]
and define $\cG^L$ as a collection of probability measures defined by such finite compositions, that is,
\[
    \cG^L
    =
    \{
        \mu(\G(\cdot)) :
        \G \in \bG^L
    \}.
\]
Hence $\cG^L$ includes all measures that can be expressed in the form of ensemble $G_1 \oplus \cdots \oplus G_K$.
\end{enumerate}

We also need to review the definition of push-forward measures because this notation is closely related to the operation of residualization.
Let $\varphi$ be a mapping $\Omega \mapsto \Omega$ and $H$ be a probability measure.
Then the push-forward of $H$ is defined in the following form:
\[
    \varphi \# H(B)
    =
    H ( \varphi^{-1}(B))
    \text{ for }
    B \in \mathcal{B}(\Omega).
\]
The following lemma establishes a connection between the ensemble measure and the push-forward measures.

\begin{lem}
\label{lem: push-forward}
For a probability measure $F$, $F \in \cG^L$ holds if and only if there exists a mapping $\G \in \bG^L$ such that $\G \# F = \mu$.
\end{lem}

\noindent
(Proof)
Suppose $F \in \cG^L$. 
Then there exists a mapping $\G \in \bG^L$ such that 
\[
    F(B) = \mu (\G(B)) \text{ for } B \in \mathcal{B}(\Omega).
\]
From Proposition \ref{prop: G is bijective and measurable}, $\G$ is bijective. 
Hence for $B \in \mathcal{B}(\Omega)$, we have
\[
    \mu(B)
    =  \mu (\G \circ \G^{-1}(B)) = 
    F(\G^{-1}(B)),
\]
so $\G \# F = \mu$.
The necessity can be shown in the same way.
\qed
\\

In the rest of the section, we first discuss the expressive power of the tree ensemble for the uni-variate cases and next generalize the result for the multi-variate cases.
After that, this result is used to prove Theorem \ref{thm: universal approximation property}.

\subsection{Uni-variate Cases}
The following proposition shows that any distribution with piece-wise constant and positive densities can be represented in the form of tree ensemble.

\begin{prop}
\label{prop: univariate}
    Let $F$ be a probability measure that admits the piece-wise constant density $f$ with the following form
    \[
        f(x) = \sum^I_{i=1} \beta_i 1_{(c_{i-1}, c_i]},
    \]
    where $\beta_i >0$ for $i =1,\dots,I$ and
    \[
        0 = c_0 < c_1 < \cdots < c_I = 1.
    \]
    Then, if $L \geq 2$, $F \in \cG^L$ holds.
\end{prop}

\noindent 
(Proof)
We first show the existence of a tree CDF $\G_1 \in \bG^2_0$ such that the push-forward measure $\G_1 \# F$ has a density $f_1$ with the following form
\begin{align}
    f_1(x) = \sum^{I-1}_{i=1} \tilde{\beta}_i 1_{(\tilde{c}_{i-1}, \tilde{c}_i]},
    \label{eq: f1}
\end{align}
where $\tilde{\beta}_i >0$ for $i=1,\dots,I-1$ and $0 = \tilde{c}_0 < \tilde{c}_1 < \cdots < \tilde{c}_{I-1} = 1$.

Let $\alpha \in (0,1)$ be a constant that satisfies
\[
    \frac{1-\alpha}{\alpha} = \frac{\beta_2}{\beta_1}
    \frac{1 - c_1}{c_1}.
\]
Then define a measure $G_1 \in \cG^2_0$ such that 
\[
    G_1((0, c_1]) = \alpha,\ 
    G_1((c_1, 1]) = 1-\alpha
\]
and $G_1$ is conditionally uniform on $(0, c_1]$ and $(c_1, 1]$.
Let $\G_1$ be $G_1$'s tree CDF and $F_1 = \G_1 \# F$ be a probability measure with the density $f_1$.
For $x \in (0,\alpha]$, we have
\begin{align*}
    F_1 ((0,x]) 
    = F(\G^{-1}_1((0,x]))
    = F((0, \G^{-1}_1(x)])
    =
    \int^{\G^{-1}_1(x)}_0 f d \mu.
\end{align*}
Hence, by the chain rule, the density at this $x$ is written as
\[
    f_1(x) = \frac{c_1}{\alpha} f \left(
        \G^{-1}_1(x)
    \right)=
    \frac{c_1}{\alpha} \beta_1
    .
\]
Similarly, the density at $x \in(\alpha, 1]$ is written as
\[
    f_1(x) =
    \frac{1-c_1}{1-\alpha}
    f \left(
        \G^{-1}_1(x)
    \right).
\]
Let $\tilde{c}_i = \G_1(c_{i+1})$ for $i=1,\dots,I-1$. 
By this definition, $\alpha < \tilde{c}_1$, and the density of $f_1$ at $x \in (\alpha, \tilde{c}_1]$ satisfies
\begin{align*}
    f_1 (x)
    =
    \frac{1-c_1}{1-\alpha}
    \beta_2=
    \frac{c_1}{\alpha} \beta_1,
\end{align*}
where the second equation follows the definition of $\alpha$.
Hence $f_1$ is constant on $(0,\tilde{c}_1]$.
Moreover, the density on $(\tilde{c}_{i-1}, \tilde{c}_{i}]$ for $i=2,\dots,I-1$ is $(1-c_1)/(1-\alpha) \beta_{i-1}$ so constant. 
Therefore the density $f_1$ is written in the form of Equation~\ref{eq: f1}.

By using the same logic for the rest of the $I-2$ discontinuous points, we can define tree CDFs $\G_2, \cdots, \G_{I-1}$ that connect the densities at these points one by one. 
Hence the measure $(\G_{I-1}\circ \cdots \circ \G_1) \# F$ has a constant density and thus is the uniform measure $\mu$.
\qed

\subsection{Multi-variate Cases}
In this section, we prove the following proposition that is a generalization of Proposition \ref{prop: univariate}.

\begin{prop}
    \label{prop: multi}
    For $j = 1,\dots,d$, let $\{c_{j, i_j}\}_{i_j=1}^{I_j}$ be a sequence such that
    \[
        0 = c_{j, 0} < c_{j, 1} < \cdots < c_{j,I_j} = 1,
    \]
    and $\mathcal{L} =  \{A_{i_1,\dots,i_d}\}_{i_1,\dots,i_d}$ be a partition of the sample space $(0,1]^d$ that consists of rectangles written as
    \[
        A_{i_1,\dots,i_d}
        =
        (c_{1, i_1-1}, c_{1, i_1}]
        \times
        \cdots 
        \times 
        (c_{d, i_d-1}, c_{d, i_d}].
    \]
    If a probability measure $F$ is piecewise uniform on $\mathcal{L}$ and written as
    \[
        F(B)
        =
        \sum_{i_1,\dots,i_j}
        a_{i_1,\dots,i_d}
        \frac{\mu(B \cap A_{i_1,\dots,i_d})}{\mu(A_{i_1,\dots,i_d})},
        \text{ for }
        B \in \mathcal{B}((0,1]^d),
    \]
    where $a_{i_1,\dots,i_d} > 0$, then for $L \geq d+1$, there is a mapping $\G \in \bG^L$ such that $\G \# F = \mu$ and thus $F \in \cG^L$.
    In addition, we can choose $\G$ so that for every pair of indices $(i_1,\dots,i_d)$, the image $\G(A_{i_1,\dots,i_d})$ is a rectangle written as
    \[
        (\G(c_{1, i_1-1}), \G(c_{1, i_1})]
        \times
        \cdots 
        \times 
        (\G(c_{d, i_d-1}), \G(c_{d, i_d})].      
    \]
\end{prop}

\noindent 
(Proof)
We use induction: We assume that the statement of Proposition \ref{prop: multi} is valid for the 1,2,\dots,(d-1)-dimensional cases.
Because in this proof we handle measures and transformation defined in different dimensional spaces, the sets $\cG^L$ and $\bG^L$ defined for the $j$-dimensional space are denoted by $\cG^{L,d}$ and $\bG^{L,d}$, respectively.

Inside of the induction, we also assume that for some $l \in \{1,\dots,I_d-1\}$, there are mappings $\G_1,\dots, \G_l \in \bG^{L,d}$ such that a probability measure $F_l := (\G_l \circ \cdots \circ \G_1) \# F$ is a piecewise uniform probability measure written as, for $B \in \mathcal{B}((0,1]^d)$,
\begin{align*}
    F_l(B)
    &=
    \sum^l_{i=1}
    C_i
    \frac{\mu(B \cap (0,1]^{d-1} \times (c_{d, i-1}, c_{d, i}])}{\mu((0,1]^{d-1} \times (c_{d, i-1}, c_{d, i}])}+
    \sum^{I_d}_{i_d=l+1}
    \sum_{i_1,\dots,i_{d-1}}
    a^{(l)}_{i_1,\dots,i_d}   
    \frac{\mu(B \cap A^{(l)}_{i_1,\dots,i_d})}{\mu(A^{(l)}_{i_1,\dots,i_d})},
\end{align*}
where $C_i > 0$ and $a^{(l)}_{i_1,\dots,i_d}>0$ for all indices. Also, for the second term, $A^{(l)}_{i_1,\dots,i_d}$ is a rectangular written as
\begin{align*}
    A^{(l)}_{i_1,\dots,i_d}
    &=
    \left(
        c^{(l)}_{1,i_1-1}, c^{(l)}_{1,i_1}
    \right] \times
    \cdots 
    \times 
    \left(
        c^{(l)}_{d-1,i_{d-1}-1},
        c^{(l)}_{d-1,i_{d-1}}
    \right]\\
    &
    \hspace{10mm}
    \times 
    (
        c_{d, I_d -1}, c_{d, I_d}
    ],
\end{align*}
where for $j=1,\dots,d-1$, $\{c^{(l)}_{j,i}\}^{I_j}_{i=1}$ is a sequence such that
\[
    0 = c^{(l)}_{j,1} < c^{(l)}_{j,2} < \cdots < c^{(l)}_{j, I_j} = 1.
\]
(We note that this sequence's length can be different from ``$I_j$'' provided in Proposition \ref{prop: multi} but to avoid an excessive number of indices, we use $I_j$ here because its size does not affect the logic provided in this proof.)
Under this assumption, we show that there is a measure (``$F_{l+1}$'') that has the same form for $l+1$.

Define a $d-1$-dimensional probability measure $\hat{F}_{l+1}$
\begin{align*}
    \hat{F}_{l+1}
    &=
    \sum_{i_1,\dots,i_{d}}
    \frac{a^{(l)}_{i_1,\dots,i_{d-1}, l+1}}{C_{l+1}}
    \frac{\mu_{d-1} (B \cap \hat{A}^{(l)}_{i_1,\dots,i_{d-1}})}{\mu_{d-1}(\hat{A}^{(l)}_{i_1,\dots,i_{d-1}})}
    \text{ for }
    B \in \mathcal{B}((0,1]^{d-1}),
\end{align*}
where $C_{l+1}$ is the normalizing constant, $\mu_{d-1}$ is the Lebesgue measure defined for the $d-1$-dimensional sample space, and $\hat{A}^{(l)}_{i_1,\dots,i_{d-1}}$ is a set written as
\begin{align}
    \hat{A}^{(l)}_{i_1,\dots,i_{d-1}}
       &=
    \left(
        c^{(l)}_{1,i_1-1}, c^{(l)}_{1,i_1}
    \right] \times
    \cdots 
    \times 
    \left(
        c^{(l)}_{d-1,i_{d-1}-1},
        c^{(l)}_{d-1,i_{d-1}}
    \right] .
    \label{def: d-1 dimensitonal rectangle}
\end{align}
Because $\hat{F}$ is a piecewise uniform measure defined on the partition that consists of hyper-rectangles, by the assumption we set for the induction, there is a mapping $\hat{\G}_{l+1} \in \bG^{L-1, d-1}$ such that $\hat{\G}_{l+1} \# \hat{F}_{l+1} = \mu_{d-1}$.
With this mapping, we define a mapping $\G_{l+1}: (0,1]^d \mapsto (0,1]^d$ such that for $x = (x_1,\dots,x_d) \in (0,1]^d$,
\[
    \G_{l+1}
    (x)
    =
    \left(
        \hat{\G}_{l+1}(x_1,\dots,x_{d-1}),
        x_d
    \right)
\]
if $x_d \in (c_{d,l}, 1]$ and otherwise $\G_{l+1}
    (x) = x$.
The mapping $\G_{l+1}$ moves points only in $(0,1]^{d-1} \times (c_{i,l},1]$, which is a node one can obtain by dividing the sample space only once, according to $\hat{\G}_{l+1}$, which is a mapping that is a composition of tree CDFs based on trees with $L-1$ leaf nodes. 
Hence $\G_{l+1}$ is a composition of tree CDFs defined on trees with $(L-1)+1 = L $ leaf nodes, so we have $\G_{l+1} \in \bG^{L,d}$.
With this mapping, we define a measure $F_{l+1} = \G_{l+1} \# F_l$.

Fix a pair of indices $(i_1,\dots,i_d)$ and let $B_{d-1} \in \mathcal{B}((0,1]^{d-1})$ and $B_1 \in \mathcal{B}((0,1])$ be measurable sets such that
\[
    B_{d-1} \times B_1
    \in \hat{\G}_{l+1} (\hat{A}^{(l)}_{i_1,\dots,i_{d-1}})
    \times
    \left(
        c_{d, i_d - 1}
        ,
        c_{d, i_d}     
    \right].
\]
If $i_d \leq l$, by the definition of $F_{l+1}$ and $\G_{l+1}$,
\[
    F_{l+1}(B_{d-1} \times B_1)
    =
    F_l (\G^{-1}_{l+1}(B_{d-1} \times B_1))
    =
    F_l (B_{d-1} \times B_1).
\]
On the other hand, if $i_d \geq l+1$, since $F_l$ is conditionally uniform on $A^{(l)}_{i_1, \dots, i_d}$,
\begin{align*}
    F_{l+1}(B_{d-1} \times B_1)
    &=
    F_l (\G^{-1}_{l+1}(B_{d-1} \times B_1))
    =
    F_l (\hat{\G}^{-1}_{l+1}(B_{d-1}) \times B_1) \\
    &=
    a^{(l)}_{i_1,\dots,i_d}
    \frac{\mu(\hat{\G}^{-1}_{l+1}(B_{d-1}) \times B_1)}
    {\mu(A^{(l)}_{i_1, \dots, i_d})}\\
    &=
    a^{(l)}_{i_1,\dots,i_d}
    \frac{\mu_{d-1}(\hat{\G}^{-1}_{l+1}(B_{d-1}))
    \mu_1(B_1)
    }
    {\mu_{d-1}(\hat{A}^{(l)}_{i_1,\dots,i_{d-1}})
    \mu_1(   \left(
        c_{d, i_d - 1}
        ,
        c_{d, i_d}     
    \right])
    },
\end{align*}
where $\mu_1$ is the Lebesgue measure defined for the 1-dimensional sample space.
For such $i_d$, by the definition of $\hat{F}_{l+1}$ and $\hat{\G}_{l+1}$,
\begin{align*}
    \mu_{d-1}(B_{d-1})
    &=
    \hat{F}_{l+1} (\hat{\G}^{-1}_{l+1} (B_{d-1}))
    =
    \frac{a^{(l)}_{i_1,\dots,i_{d-1},l+1}}{C_{l+1}}
    \frac{\mu_{d-1}
      (\hat{\G}^{-1}_{l+1}(B_{d-1}))
    }{
        \mu_{d-1}(\hat{A}^{(l)}_{i_1,\dots,i_{d-1}})
    },
\end{align*}
from which we obtain
\begin{align}
    F_{l+1}(B_{d-1}\times B_1)
    &=
    C_{l+1}
    \frac{a^{(l)}_{i_1,\dots,i_{d-1},i_d}}
    {a^{(l)}_{i_1,\dots,i_{d-1},l+1}}
    \frac{\mu_{d-1}(B_{d-1})\mu_1(B_1)}{\mu_1(   \left(
        c_{d, i_d - 1}
        ,
        c_{d, i_d}     
    \right])}
    \nonumber 
    \\
    &=
    \begin{cases}
      C_{l+1} 
         \frac{\mu(B_{d-1}\times B_1)}{\mu
         \left( 
         (0,1]^{d-1}
         \times
         (
        c_{d, i_d - 1}
        ,
        c_{d, i_d}     
    ]
    \right)}
    &
    (i_d = l+1),\\
    a^{(l+1)}_{i_1,\dots,i_d}
    \frac{\mu(B_{d-1} \times B_1)}
    {\mu( \tilde{A}^{(l+1)}_{i_1,\dots,i_d} )}
    &
    (i_d > l+1),
    \label{def of F l+1}
    \end{cases}
\end{align}
where, for $i_d > l+1$,
\begin{align*}
    \tilde{A}^{(l+1)}_{i_1,\dots,i_d} 
    &=
    \hat{\G}_{l+1}
    (\hat{A}^{(l)}_{i_1,\dots,i_d})
    \times
    (
        c_{d, i_d-1},
        c_{d,i_d}
    ].
\end{align*}
and
\begin{align*}
    a^{(l+1)}_{i_1,\dots,i_d}
    &=
    C_{l+1}
    \frac{a^{(l)}_{i_1,\dots,i_{d-1},i_d}}
    {a^{(l)}_{i_1,\dots,i_{d-1},l+1}}
    \mu_{d-1}
    \left(
        \hat{\G}_{l+1}
    (\hat{A}^{(l)}_{i_1,\dots,i_d})
    \right).
\end{align*}
Because $\mathcal{B}((0,1]^{d-1}) \times\mathcal{B}((0,1])$ generates $\mathcal{B}((0,1]^{d-1})$, from the discussion provided above, $F_{l+1}$ is piecewise-uniform on a partition that consists of
\[
    (0,1]^{d-1} \times \left(
        c_{d, i_d - 1}
        ,
        c_{d, i_d}     
    \right]\ (i_d \leq l+1)
\]
and $\tilde{A}^{(l+1)}_{i_1,\dots,i_d} $ ($i_d > l+1$), and this partition is denoted by $\mathcal{P}^{(l+1)}$.
Note that by the definition of $\hat{\G}_{l+1}$, and Proposition \ref{prop: multi}, which we assume holds for $(d-1)$-dimensional cases, the image of the hyper-rectangle of $\hat{A}^{(l)}_{i_1,\dots,i_{d-1}}$ under $\hat{\G}_{l+1}$ and $\tilde{A}^{(l+1)}_{i_1,\dots,i_d} $ are a hyper-rectangle in the $(d-1)$-dimensional space and the $d$-dimensional space, respectively.

The following lemma states that the partition structure $\mathcal{P}^{(l+1)}$ has a finer partition that has the ``checker-board'' form, as shown in the next lemma.

\begin{lem}
\label{lem on partition with rectanbles}
Let $\{D_i\}^I_{i=1}$ is a partition of the sample space $(0,1]^d$ such that every $D_i$ is a hyper-rectangle. 
Then, there are sequences $\{e_{j, i_j}\}_{i_j}$ ($j=1,2,\dots,d$) such that
\[
    0 = e_{j, 0} < e_{j, 1} < \cdots < e_{j,I_j} = 1
\]
and a partition $\{E_{i_1,\dots,i_d}\}_{i_1,\dots,i_d}$ defined as
\[
    E_{i_1,\dots,i_d}
    =
    (e_{1, i_1-1}, e_{1, i_1}]
    \times
    \cdots 
    \times 
    (e_{d, i_d-1}, e_{d, i_d}]
\]
such that every $D_i$ is a finite union of elements of $\{E_{i_1,\dots,i_d}\}_{i_1,\dots,i_d}$.
\end{lem}

Its proof is straightforward because we only need to ``extend'' the boundaries between the rectangles $\{D_i\}^I_{i=1}$.
By applying this lemma to the partition $\mathcal{P}^{(l+1)}$, it follows that there are finite sequences $\{c^{(l+1)}_{i_j}\}_{i_j}$ ($j=1,\dots,d-1$) such that a checkerboard-like partition consisting of the following type of rectangles
\begin{align*}
    A^{(l+1)}_{i_1,\dots,i_d}
    :=&(
        c^{(l+1)}_{1, i_1-1},
        c^{(l+1)}_{1, i_1}
    ]
    \times
    \cdots
    \times
    (
        c^{(l+1)}_{d-1, i_{d-1}-1},
        c^{(l+1)}_{d-1, i_{d-1}}
    ]
    \times
    (
        c_{d, i_d-1},
        c_{d, i_d}
    ]    
\end{align*}
is finer than $\mathcal{P}^{(l+1)}$.
With this partition, the measure $F_{l+1}$ is written as
for $B \in \mathcal{B}((0,1]^d)$
\begin{align*}
    F_{l+1}(B)
    &=
    \sum^{l+1}_{i=1}
    C_i
    \frac{\mu(B \cap (0,1]^{d-1} \times (c_{d, i-1}, c_{d, i}])}{\mu((0,1]^{d-1} \times (c_{d, i-1}, c_{d, i}])}+
    \sum^{I_d}_{i=l+2}
    \sum_{i_1,\dots,i_{d-1}}
    a^{(l)}_{i_1,\dots,i_d}   
    \frac{\mu(B \cap A^{(l)}_{i_1,\dots,i_d})}{\mu(A^{(l)}_{i_1,\dots,i_d})}.
\end{align*}
Because this result holds for $l=1,\dots,I_d-1$, there exists a sequence of mappings \\ $\G_1,\dots,\G_{I_d} \in \bG^{L,d}$ such that a push-forward measure $H := (\G_{I_d} \circ \dots \circ \G_1)\# F$ has a form 
\[
    H(B)
    =
    \sum^{I_d}_{i=1}
    C_i
    \frac{\mu(B \cap (0,1]^{d-1} \times (c_{d, i-1}, c_{d, i}])}{\mu((0,1]^{d-1} \times (c_{d, i-1}, c_{d, i}])}.
\]

Define an one-dimensional probability measure $\hat{H}$ as follows
\[
    \hat{H}(B_1)
    =
    \sum^{I_d}_{i=1}
    C_i
    \frac{\mu_1(B_1 \cap (c_{d, i-1}, c_{d, i}])}{\mu_1((c_{d, i-1}, c_{d, i}])}
    \text{ for }
    B_1 \in \mathcal{B}((0,1]).
\]
Then, by the assumption (or Proposition \ref{prop: univariate}), there exists a mapping $\hat{G}_0 \in \bG^{2,1}$ such that $\hat{\G}_0 \# \hat{H} = \mu_1$.
With this mapping, we define another mapping $\G_0: (0,1] \mapsto (0,1]$ such that for $x = (x_1,\dots,x_d)$,
\[
    \G_0(x)
    =
    (x_1,\dots,x_{d-1}, \hat{G}_0(x_d)).
\]
This mapping moves input points only in the $d$th dimension according to $\hat{G}_0$ so it is written as a composition of tree CDFs defined on trees with $2$ terminal node and thus an element of $\mathcal{G}^{L,d}$.
Hence $\G_0 \in \bG^{L,d}$.
Fix $i \in \{1,\dots,I_d \}$.
For a measurable set $B_{d-1} \times B_1$ such that
\[
    B_{d-1} \times B_1
    \in
    \mathcal{B}((0,1]^{d-1})
    \times 
    \hat{G}_0 (
     (c_{d,i-1}, c_{d,i}]
    ),
\]
because $H$ is piecewise uniform, we have
\begin{align*}
    \G_0 \#  H
    (B_{d-1} \times B_1)
    &=
    H(\G^{-1}_0 (B_{d-1} \times B_1))
    =
    H(B_{d-1} \times \hat{\G}^{-1}_0 (B_1))\\
    &=
    C_i
    \frac{
    \mu(B_{d-1} \times \hat{\G}^{-1}_0(B_1))
    }{\mu((0,1]^{d-1} \times (c_{d, i-1}, c_{d, i}])} \\
    &=
    C_i
    \frac{\mu_{d-1}(B_{d-1}) \mu_1(\hat{\G}^{-1}_0 (B_1))}
    {\mu_1 ((c_{d, i-1}, c_{d, i}]) }.
\end{align*}
On the other hand,
\[
    \mu_1 (B_1)
    =
    \hat{H} (\hat{\G}^{-1}_0(B_1))
    =
    C_1
    \frac{\mu_1(\hat{\G}^{-1}_0(B_1))}
    {\mu_1 ((c_{d, i-1}, c_{d, i}]) }.
\]
Hence, we obtain
\[
    \G_0 \#  H
    (B_{d-1} \times B_1)
    =
    \mu_{d-1}(B_{d-1})
    \mu_1(B_1)
    =
    \mu(B_{d-1} \times B_1).
\]
Therefore, we conclude that
\[
\G_0 \#  H
=
(\G_0 \circ \G_{I_d} \circ \cdots \circ \G_1)\#H
=
\mu. \qed
\]

The result of Proposition \ref{prop: multi} can be described in a simplified form as in the next corollary.
This proof immediately follows Proposition \ref{prop: multi} and Lemma \ref{lem on partition with rectanbles}.

\begin{cor}
    \label{cor: multi-general}
    Let $\{E_i\}^I_{i=1}$ is a partition of the sample space $(0,1]^d$ such that $E_i$ is a rectangle with a form
    \[
        E_i 
        =
        (a_{i,1}, b_{i,1}] \times \cdots \times (a_{i,d}, b_{i,d}],
    \]
    and $F$ be a piecewise uniform probability measure defined on the partition: 
    \[
        F(B)
        =
        \sum_{i=1}^I 
        \beta_i \frac{\mu(B \cap E_i)}{\mu(E_i)}
        \text{ for }
        B \in \mathcal{B}((0,1]^d),
    \]
    where $\beta_i > 0$. 
    Then $F \in \cG^L$ for $L \geq d+1$.
\end{cor}

\subsection{Proof of Theorem \ref{thm: universal approximation property}}
We finally provide the proof of Theorem \ref{thm: universal approximation property}, which can be obtained by adding minor modifications to the proof of Theorem 4 in \cite{wong2010optional}.

Let $f^*$ denote $F^*$'s density function, and we first assume that $f^*$ is uniformly continuous.
For $\epsilon>0$, there exists $\epsilon' > 0$ such that $\log(1+\epsilon') < \epsilon$. 
Since the function $f^*$ is uniformly continuous, there exists $\delta > 0$ such that
\[
    |x - y| < \delta
    \Rightarrow 
    |f^*(x) - f^*(y)| < \epsilon'.
\] 
Let $\{E_i\}^I_{i=1}$ is a partition of the sample space $(0,1]^d$ such that $E_i$ has a rectangle shape and $\mathrm{diam}(E_i) < \delta$. 
Define a function $\tilde{g}$ as
\[
    \tilde{g}
    =
    \sum^I_{i=1}
    \left\{
    \sup_{x \in E_i} f^*(x)
    \right\}
    \1_{E_i}(x)
    \text{ for }
    x \in (0,1]^d.
\]
Let $C = \int \tilde{g} d\mu$.
Because $\tilde{g}(x) \geq f^*(x)$ for $x \in (0,1]^d$, we have $C \geq 1$ and 
\begin{align*}
    0 &\leq C -1
    = \int (\tilde{g} - f^*) d\mu
    =
    \sum^I_{i=1}
    \int_{E_i}
    (\tilde{g}(x) - f^*(x)) d \mu \\
    &\leq 
    \sum^I_{i=1} \int_{E_i} \epsilon' d\mu 
    = \epsilon'.
\end{align*}
Define a density function $g:= \tilde{g} / C$.
The corresponding probability measure $G$ is an element of $\cG^{L}$ by Corollary \ref{cor: multi-general}.
Hence, for the two measures $F^*$ and $G$, we can bound the KL divergence as follows
\begin{align*}
    KL(F || G)
    &=
    \int f^* \log \frac{f^*}{g} d\mu
    =
    \int f^* \log \frac{f^*}{\tilde{g}} d\mu 
    +
    \int f^* \log C d\mu \\
    &\leq 
    \log C \leq \log(1+ \epsilon') < \epsilon. 
\end{align*}
We next consider the general case, where we assume $f^* \leq M$ for some $M >0$. 
By Lusin's theorem, for any $\tilde{\epsilon} > 0$, there exits a closed set $B$ such that $\mu(B^c) < \tilde{\epsilon}$ and $f^*$ is uniformly continuous on $B$.
Using this fact, we modify the first discussion as follows.
The definition of $\tilde{g}$ is modified as follows: If $E_i \cap B \neq \emptyset$, for $x \in E_i$, we let
\[
    \tilde{g}(x) = \sup_{x \in E_i \cap B} f^*(x).
\]
Otherwise, $g(x) = M$.
With this modification, we obtain
\begin{align*}
0 \leq C-1
&= \int (\tilde{g} - f^*) d\mu 
=
\int_B (\tilde{g} - f^*) d\mu
+
\int_{B^c} (\tilde{g} - f^*) d\mu\\
&\leq 
\epsilon' + M \tilde{\epsilon},
\end{align*}
which can be arbitrarily small, so the same result follows.
\qed

\section{Details on Learning Probability Measures with the \Polya Tree Process}
\label{supp: PT based sampling}

This section provides details on the weak learner we use
to fit tree measures to the residuals in the estimation.
The algorithm is based on the PT-based method proposed in \cite{awaya2022hidden}, and interested readers may refer to this paper.

\subsection{Theoretical Justification of Using the PT-based Model}
As shown in Section Proposition~\ref{prop: optimal solution to minimize KL (finite)}, the improvement in the entropy loss is maximized when a fitted tree is a solution of the problem
\begin{align*}
    \argmax_{T \in \T} 
    \sum_{A \in \mathcal{L}(T)}
    \tilde{F}^{(n)}_k (A)
    \log \frac{\tilde{F}^{(n)}_k(A)}{\mu(A)},
\end{align*}
where $\tilde{F}^{(n)}_k$ is the empirical measure defined by the residuals ${\bm r}^{(k-1)}=\{r^{(k-1)}_i\}_{i=1}^{n}$. 
As $n \to \infty$, the empirical measure $\tilde{F}^{(n)}_k (B)$ converges to  $\tilde{F}_k (B)$ for $B \subset (0,1]^d$, where $\tilde{F}_k$ is the true distribution of the residuals defined by the previous tree-CDFs $\G_1,\dots,\G_{k-1}$.
At this population level, the maximization problem is written as
\begin{align*}
    \argmax_{T \in \T} 
    \sum_{A \in \mathcal{L}(T)}
    \tilde{F}_k (A)
    \log \frac{\tilde{F}_k(A)}{\mu(A)},
\end{align*}
and we can show that this maximization is equivalent to minimizing the KL divergence $KL(\tilde{F}_k || \tilde{F}_k |_T)$, where $\tilde{F}_k |_T$ is ``a tree-approximation of $\tilde{F}_k$ under $T$'', namely,
\[
    \tilde{F}_k |_T (B)
    =
    \sum_{A \in \mathcal{L}(T)}
    \tilde{F}_k (A)
    \frac{\mu(B \cap A)}{\mu(A)} \text{ for } B \in (0,1]^d.
\]
Theorem 4.1 in \cite{awaya2022hidden} shows that the posterior of trees also concentrates on the minimizer of $KL(\tilde{F}_k || \tilde{F}_k |_T)$, and this result implies that at the population level, or when $n$ is large, we can find the tree that maximizes the improvement in the entropy loss or similar ones by checking the posterior of trees.

\subsection{Details on the sampling algorithm}

Suppose we have obtained the residuals at the beginning of the boosting algorithm.
Since the task of fitting a new measure to the residuals is essentially the same for all steps, we drop the $k$, the index of the trees and measures consisting of the ensemble, from the notations for simplicity.
Then the residuals are denoted by $\bm{r} = (r_1,\dots,r_n)$, and our task at each step is to capture their distributional structure by fitting a dyadic tree.
In the section, we provide details on the prior distributions introduced for the tree $T$ and the stochastic top-down algorithm we use to find a tree with good fitting.

\subsubsection{Prior distribution of $T$}

As in \cite{awaya2022hidden}, we construct a prior of $T$ by introducing the random splitting rule for each node $A$.
First, we introduce the stopping variable $S(A)$ that takes 0 or 1, and if $S(A) = 1$, we stop splitting $A$ and otherwise split $A$.
Here we set $P(S(A) = 1)$ to $0.5$.
In the latter case, we next define the dimension variable $D(A)$ and the location variable $L(A)$.
If $D(A) = j$ ($j=1,\dots,d$), the node $A$ is split in the $j$th dimension, and the location of the boundary is determined by $L(A) \in (0,1)$, in which 0 (or 1) corresponds to the left (or right) end point.
Their prior distributions are as follows: 
\begin{align*}
P(D(A) = j ) &= 1/d, \  (j = 1,\dots,d),\\
P(L(A) = l/N_L) &= \frac{1}{N_L - 1} \  (l = 1,\dots,N_L-1),
\end{align*}
where $N_L - 1$ is the number of grid points, which is 127 in the estimation.

On the tree $T$, we also define a random measure $\tilde{G}$, with which we can define the likelihood of the residuals $\bm{r}$.
The prior of the measure $\tilde{G}$ is defined by introducing the parameters $\theta(A) = \tilde{G}(A_l \mid A)$, where $A_l$ is the left children node, for every non-terminal node $A$.
They follow the prior distribution specified as
\[
    \theta(A) 
    \sim 
    Beta (\theta_0(A), 1-\theta_0(A)),\ 
    \theta_0(A) = \frac{\mu(A_l)}{\mu(A)}.
\]
We note that this random measure $\tilde{G}$ is introduced just to define the marginal posterior of $T$, namely, $P(T \mid \bm{r})$, since our main goal is to find a tree that fits the distribution of the residuals.
In the estimation, we first choose one tree according to this posterior and construct the measure to output, which is denoted by $G_k$ in the paper, as in Equation~\ref{eq: Gk with shrinkage}.
The method to select one tree is described in more detail in the next section.

The joint model of the tree $T$ and the measure $\tilde{G}$ can be seen as a special case of the density estimation model that is referred to as the adaptive \Polya tree \citep{ma2017adaptive} model in  \cite{awaya2022hidden} with the number of the latent states being 2.

\subsubsection{Top-down stochastic algorithm}
The particle filter proposed in \cite{awaya2022hidden} is shown to be effective to sample from the posterior of trees.
This original algorithm, however, has drawbacks when seen as a component of the boosting from a viewpoint of computational cost:

\begin{enumerate}
    \item In the particle filter, we construct thousands of candidate trees, but this strategy may make the whole boosting algorithm too time-consuming since in the boosting algorithm we need to repeat fitting trees to the residuals many times.
    \item In the original algorithm, we do not stop splitting nodes until we reach the bottom nodes unless the number of included observations is too small.
    (Technically speaking, this is because the stopping variables, or the latent variables in general, are integrated out in the sampling.)
    The number of nodes generated in a tree, however, tends to be large especially when the sample size is large, and constructing such large trees repeatedly in the boosting algorithm is also too-time consuming.
    The computation cost would become reasonable if we ``give up non-promising nodes'', that is, stop dividing nodes if no interesting structures are found there.
\end{enumerate}

From these reasons, we modify the original algorithm as follows: (i) Instead of generating many candidate trees, we set the number of particles to one, that is to say, construct a tree by randomly splitting nodes on the tree in a top-down manner.
Hence the algorithm is similar to the top-down greedy method, but in our algorithm one selects splitting rules stochastically.
(ii) For each active node, we compare possible splitting rules and the decision of stopping the splitting, where the latter option is added to the algorithm. 
This comparison is based on their posterior probabilities, and the splitting tends to be stopped if the conditional distribution is close to uniform.

For an active node $A$, the possible decisions are compared based on the following quantities that are seen as ``prior $\times$ marginal likelihood''.
A conceptually very similar algorithm for supervised learning is proposed in \cite{he2021stochastic}.
For the decision of stopping, we compute
\[
    L_\emptyset
    =
    P(S(A) = 1)
    \mu(A)^{-n(A)},
\]
where $n(A)$ is the number of residuals included in $A$.
On the other hand, for the splitting rule $D(A) = j$ and $L(A) = l / N_L$ that divides $A$ into $A_l$ and $A_r$, we compute
\begin{align*}
    L_{j,l}
    &=
    P(S(A) = 0, D(A) = j, L(A) = l/N_L) \\
    &\hspace{10mm}
    \times \int Beta(\theta \mid \theta_0(A), 1 - \theta_0(A))
    \theta^{n(A_l)}
    (1-\theta)^{n(A_r) } 
    d \theta
    \\
    &\hspace{10mm} \times \mu(A_l)^{-n(A_l)} \mu(A_r)^{-n(A_r)} \\
    &= P(S(A) = 0, D(A) = j, L(A) = l/N_L) \\
    &\hspace{10mm} \times \frac{Be(\theta_0(A) + n(A_l), 1-\theta_0(A) + n(A_r))}
    {Be(\theta_0(A), 1-\theta_0(A))} \\
    & \hspace{10mm} \times \mu(A_l)^{-n(A_l)} \mu(A_r)^{-n(A_r)},
\end{align*}
where $Be(\cdot)$ is the beta function.
Based on these quantities, we choose to stop the splitting with probability 
\[
    \frac{L_\emptyset}
    {L_\emptyset
        +
    \sum^{d}_{j=1}
    \sum^{N_L-1}_{l=1}
    L_{j,l}.
    }
\]
Otherwise, we choose the splitting rule $D(A) = j$ and $L(A) = l / N_L$ with probability
\[
    \frac{L_{j,l}}
    {\sum^{d}_{j'=1}
    \sum^{N_L-1}_{l'=1}
    L_{j',l'}}.
\]

\section{Details of the 48-dimensional Experiments}
\label{sec: Details of the 48-dimensional Experiments}
In the experiment, we used the following three scenarios.

\begin{description}
\item[\it Scenario A:]$n = 1,000$, and
    \begin{align*}
        \mathrm{Normal}(\boldsymbol{\mu}, \Sigma),
    \end{align*}
    where
    \begin{align*}
        \boldsymbol{\mu}
        =
        \left[ 
            \begin{array}{c}
                1/2 \\ 
                1/2 
            \end{array}
        \right], \
        \Sigma
        =
        \left[ 
            \begin{array}{cc}
                1/8^2 & 0.95/8^2\\ 
                0.95/8^2 & 1/8^2
            \end{array}
        \right].
    \end{align*}
    \item[\it Scenario B:]$n = 5,000$, and
    \begin{align*}
        &\frac{1}{10} {\rm Beta}(x_1 \mid 1,1) \times {\rm Beta}(x_2 \mid 1,1)
        + \frac{3}{10} {\rm Beta}(x_1 \mid 15,45)\times {\rm Beta}(x_2\mid 15,45) \\
        &\hspace{2.5mm}+ \frac{3}{10} {\rm Beta}(x_1 \mid 45,15)\times {\rm Beta}(x_2 \mid 22.5,37.5)\\
        &\hspace{2.5mm}+ \frac{3}{10} {\rm Beta}(x_1 \mid 37.5,22.5)\times {\rm Beta}(x_2 \mid 45,15).
    \end{align*}
\item[\it Scenario C:]$n = 2,000$, and
    \begin{align*}
        &\frac{1}{3} \1_{[0.1, 0.45] \times [0.35, 0.9]}(x_1,x_2)
        + \frac{1}{3} \1_{[0.2, 0.8] \times [0.45, 0.5]}(x_1,x_2)\\
        &\hspace{5mm} + \frac{1}{3} \1_{[0.7, 0.9] \times [0.05, 0.6]}(x_1,x_2).
    \end{align*}
\end{description}

\section{Additional Tables and Figures}
\label{supp: tables and figures}


\begin{table}[h]
\centering
\begin{tabular}{ccccc}
$(c_0, \gamma)$ & POWER     & GAS     & HEPMASS & MINIBOONE \\ \hline
$(0.1, 0.0)$ & 0.004 & 0.060 & 0.028 & 0.075    \\
$(0.1, 0.5)$ & 0.004 & 0.023 & 0.032 & 0.064    \\
$(c_0, \gamma)$ & AReM     & CASP     & BANK &  \\ \hline
$(0.1, 0.0)$  & 0.022 & 0.038 & 0.078 &    \\
$(0.1, 0.5)$ & 0.015 & 0.030 & 0.023 &
\end{tabular}
\caption{
{
The standard deviations of the average predictive scores based on 30 different random seeds.
}
}
\label{pred scores}
\end{table}

\begin{table}[h]
\centering
\begin{tabular}{cccc}
POWER     & GAS     & HEPMASS & MINIBOONE \\ \hline
4.5 & 9.6 & 24.3 & 15.3   \\
AReM     & CASP     & BANK &  \\ \hline
1.4 & 1.6 & 9.0 &  
\end{tabular}
\caption{
{
    The average computation time (seconds) for simulating 10,000 observations based on 30 different random seeds.
    The tuning parameters $c_0$, $\gamma$ are set to $0.1$ and $0.5$, respectively.
    }
}
\label{pred scores}
\end{table}

\begin{figure}[h]
\centering
\begin{tabular}{cc}
    \includegraphics[height=7.0cm]{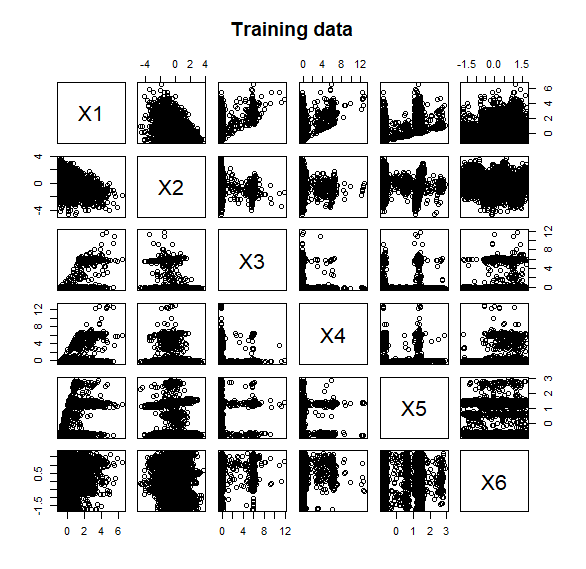}
    &
    \includegraphics[height=7.0cm]{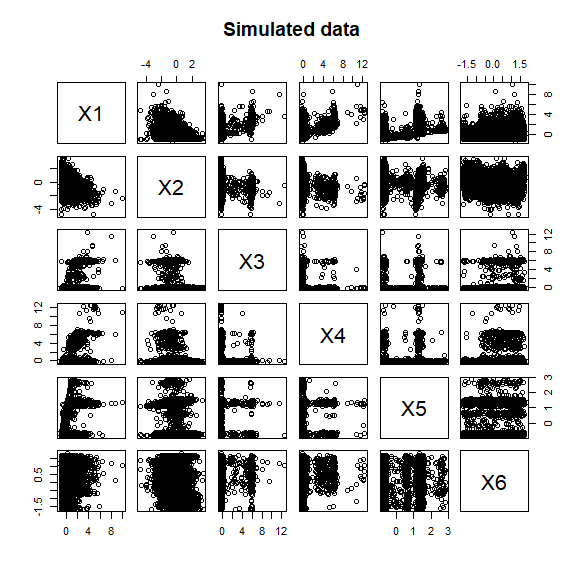}   
\end{tabular}
\caption{
    The training set of the POWER data (a subset of size 10,000 is visualized) and 10,000 observations simulated from the learned probability measure. 
}
\label{fig: train and simulation (power)}
\end{figure}

\begin{figure}[h]
\centering
\begin{tabular}{cc}
    \includegraphics[height=7.0cm]{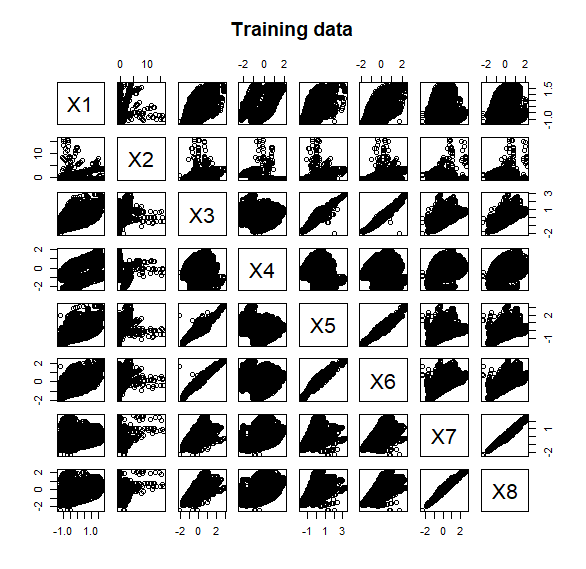}
    &
    \includegraphics[height=7.0cm]{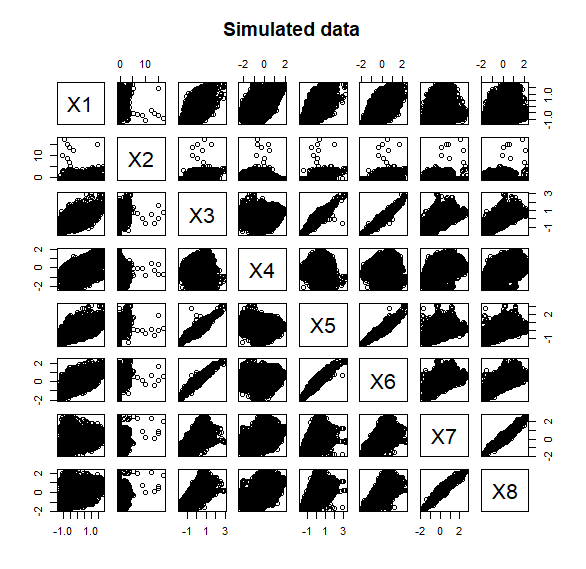}   
\end{tabular}
\caption{
    The training set of the GAS data (a subset of size 10,000 is visualized) and 10,000 observations simulated from the learned probability measure. 
}
\label{fig: train and simulation (gas)}
\end{figure}

\begin{figure}[h]
\centering
\begin{tabular}{cc}
    \includegraphics[height=7.0cm]{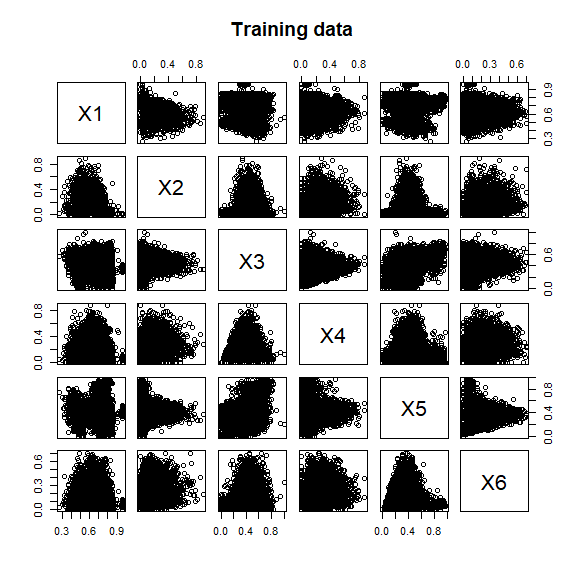}
    &
    \includegraphics[height=7.0cm]{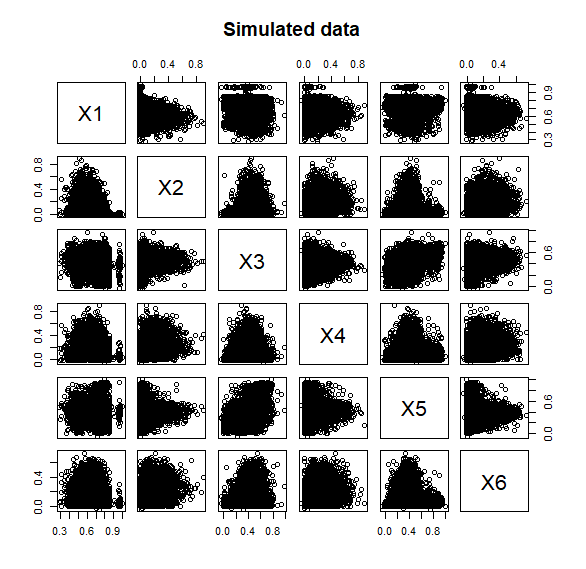}   
\end{tabular}
\caption{
    The training set of the AReM data (a subset of size 10,000 is visualized) and 10,000 observations simulated from the learned probability measure. 
}
\label{fig: train and simulation (AReM)}
\end{figure}

\end{document}